\documentclass{article}
\usepackage{graphicx} 
\usepackage{amsthm}
\usepackage{amsmath}
\usepackage{amssymb}
\usepackage{multicol}
\usepackage{pgffor}
\usepackage{caption}
\usepackage{afterpage}
\usepackage{float}
\usepackage{appendix}
\usepackage{subcaption}
\usepackage{algorithm}
\usepackage{algpseudocode}
\usepackage{hyperref}

\usepackage{graphicx} 
\usepackage{amsmath}

\usepackage[utf8]{inputenc}
\usepackage{amsmath}
\usepackage{amsthm}
\usepackage{amsfonts}
\usepackage{amssymb}
\usepackage{enumerate}
\usepackage{graphicx}
\usepackage{todonotes}
\usepackage{hyperref}

\graphicspath{ {./images/} }

\usepackage[
backend=biber,
sorting=nty
]{biblatex}
\DeclareUnicodeCharacter{03B2}{$\beta$}

\title{Numerical simulations of the spread from the mean of the SLE and Multiple SLE 
 dynamics}
\author{Phillip Kim\footnote{Georgia Institute of Technology, pkim97@gatech.edu} and Vlad Margarint\footnote{University of North Carolina at Charlotte, vmargari@charlotte.edu}}
\date{September 2024}

\begin{document}

\maketitle
\begin{abstract}
    The Schramm-Loewner Evolution (SLE) describes a family of fractal curves that arise in the study of the scaling limits of many planar Statistical Physics models. These curves are modeled using the Loewner Differential Equation for the conformal maps $g_t(z)$ with a Brownian motion driver. Using Euler's Method, in the current work we performed numerical experiments to study at a fixed time the quantities $|g_t(z) - \overline{g_t(z)}|$ and $Re(g_t(z)) - Re(\overline{g_t(z)})$, where $Re$ denotes the real part and $\overline{g_t(z)}$ refers to the sample average. These random variables measure the 'spread' of the dynamics from the average behavior at fixed time. One of the scopes of this work is to give numerical predictions for future theoretical investigations on these quantities. When investigating these quantities in the SLE case our experiments predict that the distribution is bimodal when the dynamics started close to the origin, and it can become bell-shaped if the dynamics is started further from the origin. In the second part, we performed experiments for a Multiple SLE model whose driver is Dyson Brownian Motion. Due to singularity in the dynamics of the drivers and the many data points needed, this part is challenging from a computational perspective. In the multiple SLE case, our experiments predict that the distribution is bell-shaped in all cases. In addition, we check the changes in the distributions as we vary the parameter $\kappa$ in the SLE case and $\beta$ in the Multiple SLE case. 
\end{abstract}

\section{Introduction}
The Schramm-Loewner Evolution (SLE) is a family of fractal curves parameterized by $\kappa \geq 0$ introduced by Schramm \cite{OGS} which appears in the context of studying scaling limits of many planar Statistical Physics models such as the Loop-Erased Random Walk ($\kappa = 2$) \cite{k2}, Ising Model Interfaces ($\kappa = 3$) \cite{k3}, and critical percolation on the triangular lattice ($\kappa = 6$) \cite{k6} in addition to others. 

One starting point of the SLE theory is Loewner's Differential Equation. Given a simply-connected domain $D \subset \mathbb{C}$, and a simple curve, $\gamma(0, \infty)$, then for any $t \geq 0$ there exists a conformal isomorphism from $D \setminus \gamma(0, t] \to D$ by the Riemann Mapping Theorem. We will focus on the upper half-plane, that is $D = \mathbb{H} = \{z = x + yi \in \mathbb{C} | y > 0\}$ In particular, under certain regularizations, this map, say $g_t$ satisfies the Loewner Differential Equation \cite{notes}.
\begin{equation}
    \partial_t g_t(z) = \frac{2}{g_t(z) - W_t},
\end{equation}
where $W_t$ is called the driver function. It should also be noted that this generalizes to locally growing hulls (which are compact subsets of the upper-half plane with a simply-connected complement) in place of the simple curve (see \cite{notes}). In the case of the SLE dynamics the driver is $W_t = \sqrt{\kappa}B_t$ where $\kappa \geq 0$ is a real parameter and $B_t$ is a standard one-dimensional Brownian Motion. 

In recent years, much attention was given to a multi-driver version of the Leowner Differential Equation, which is the Multiple SLE in which the Brownian Motion driver is replaced by a Dyson Brownian Motion (DBM). This DBM is a system of interacting particles. The DBM particles interact via a Coulombic repulsion, along with an independent Brownian Motion noise for each particle. For recent work on these topics, see \cite{Viv}, \cite{JiamVlad}, \cite{EveViv}, \cite{KAVlad}, \cite{Toma}  etc. In addition, we refer the reader to \cite{Zhan}, \cite{Peltola}, \cite{Dub} for other models.
This Multiple SLE model has connections with other areas such as Conformal Field Theory (see for example \cite{Cardy}).

The Dyson Brownian Motion (DBM) is described by
\begin{equation}
    d \lambda_t^{i} = \frac{\sqrt{2}}{\sqrt{N\beta}}d{B_t^i} + \frac{1}{N} \sum_{i \neq j}^{N} \frac{dt}{{\lambda_t}^i-{\lambda_t}^j}, \quad  i = 1, \dots, N
\end{equation}
and the associated Multiple Loewner Differential Equation will be
\begin{equation}
    \partial_{t} g_{t}(z) = \frac{1}{N}\sum_{i=1}^{N}\frac{2}{g_{t}(z)-\lambda_t^i}, \quad g_0(z) = z
\end{equation}
The parameters $\beta$ in the equations above and $\kappa$ from the SLE maps are linked via a simple relation $\beta=\frac{8}{\kappa}$ (see \cite{Hotta_2018}). 
DBM appeared also as a very useful tool in the modern study of Random Matrix Theory (see, for example \cite{DBM}). One of the features that was used in that direction was that DBM converges to a notion of local equilibrium faster, that is in time of order $1/N$, where $N$ is the number of particles. The global equilibrium for this particle system is achieved in a longer time of order $O(1).$ This fast convergence to local equilibrium was used in various proofs in Random Matrix Theory (see \cite{DRMT}).

One of the motivations to consider this Multiple SLE model for short times is given by the following research direction: 
We aim to investigate on the long-run what impact does the fast convergence to local equilibrium of the DBM driver has on the Multiple SLE dynamics. We hope that this initial predictions will help in the future when investigating the same or similar observable in more complicated settings related with this research direction- for example, coupling two Multiple SLE with two same-noise DBM drivers starting from two different initial conditions (see \cite{DRMT} for instances where these ideas were applied in Random Matrix Theory). 
In addition, another motivation for our work is coming from \cite{BENDER20081022} where it was shown that an analogous quantity for Dyson Brownian Motion converges to a Gaussian Process. In our current work, we wanted to explore what is the effect of the Loewner dynamics on this statistics at the level of drivers.

There are previous works on the simulation of SLE curves (see, for example,\cite{TomKen}, \cite{JamesVlad}, \cite{NV}). In our study, we dealt with the SLE maps $g_t(z)$ generated by the Loewner Differential equation. However, we do have computational challenges in simulating the $g_t(z)$ maps due to the singularity at the level of the Dyson Brownian Motion driver in the Multiple SLE case. 

In our first experiment the driver is a standard one-dimensional Brownian motion and in the second one the driver is the DBM. In both cases we consider the dynamics starting outside of a region (such that our dynamics remain in the complex plane). In the second case, the dynamics has challenges in terms of simulation as there is a singularity which appears at the level of the driver of the dynamics as the Dyson Brownian motion has a Coulomb-type interaction.  In addition, as we work with many Dyson particles, there is also a challenge of storing the data from this dynamics in order to generate the histograms from the paper. We will touch on additional challenges in the following sections.

In this manuscript, we focus on the study of the quantities $|g_t(z) - \overline{g_t(z)}|$ and $Re(g_t(z)) - \overline{Re(g_t(z)}$, where $\overline{g_t(z)}$ is the sample mean. These quantities measure the spread of the dynamics of the SLE maps from the average dynamics. We focus our simulations on two types of starting points, the ones that are close to the maximum height that Loewner hulls can reach up to in time $t \in [0, T]$ and ones that are far from this maximum height that the hull can grow up to time $T.$ We picked two representatives of points in these classes and we ran the dynamics started from them, that is we considered the SLE or the Multiple SLE dynamics started from $z_0=1.02i$ and from $z_0=3i$. The motivation behind choosing this type of scenarios is that the conformal maps tend to be more close to identity as $|z| \to \infty$ while there is more distortion captured by the maps near growth points of the Loewner dynamics. We performed the experiments for these choices of the starting points $z_0$ but the experiments with our code can be redone for all other starting points.

Another aspect that we were interested in this study is the dependence on the parameter $\beta$ of the quantities that measure the spread of the dynamics. This problem is linked with the famous still open problem of the continuity in $\kappa$ problem for the SLE curves and related objects (see \cite{Vikcont}, \cite{FrizYuan}, \cite{Vladweld}, \cite{Vladcont} for recent progress in this direction). We performed experiments for various choices of $\kappa$ for the SLE case and $\beta$ for the Multiple SLE case to investigate how these observables depend on these parameters. 

Given the importance of the SLE and Multiple SLE dynamics for the Statistical Physics and Conformal Field Theory literature, and its computation challenges we hope that this project will generate interest in the Scientific Computing community to create tools, for addressing the various challenges of the Multiple SLE dynamics such as the singularity of the drivers, the great number of data points needed to store the data, etc. This project represents our first attempt to model this dynamics and its spread around the average. We hope in future work will improve the numerical schemes (such as the use Tamed Euler Schemes for simulating DBM, see \cite{Helen}) in order to study finer properties of these dynamics.

The paper is organized as follows: After the Introduction, in Section 2, we describe the Numerical method used, in Section 3 we present part of our results from the numerical simulation for the SLE dynamics and in Section 4 a selection of results in the Multiple SLE case. Section 5 is focused on Future Directions. In the Appendix, we included outputs of the experiments of the simulations for various choices of the parameter $\beta$ and $\kappa$ for both the SLE and the Multiple SLE cases.

\section{Numerical Methods}
\begin{figure}[H]
    \centering
    \includegraphics[scale=0.4]{
        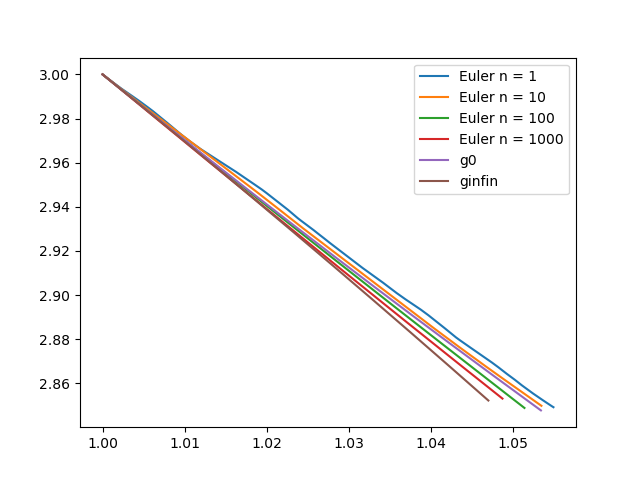
        }
    \caption{The plot of $g_t(z)$ starting from $1 + 3i$. In this figure, ginfin which is the theoretical map as $N \to \infty$ and g0 which is the map for when $\kappa = 0$ are plotted for reference. We observe that as N increases, our simulations indicate the dynamics is approaching the theoretical prediction.}
    \label{fig:EST}
\end{figure}

Our simulations used an Euler Method for both the Dyson Brownian Motions and for simulating the trajectory of the SLE dynamics driven by the Dyson Brownian motion (for the general $N$ case), or Brownian motion (for the case $N=1$). In our experiments, $N$ will take various values as indicated by the captions of the images. Given a number $N>1$, a fixed final time $T > 0$, and a fixed time step $\Delta t$, our DBM will be simulated using
\begin{equation}
    \lambda_{t_{j+1}}^{i} = \lambda^i_{t_{j}} + \frac{\sqrt{2}}{\sqrt{N\beta}}\Delta B_j^i + \frac{1}{N} \sum_{i \neq k}^{N} \frac{\Delta t}{{\lambda_{t_j}}^i-{\lambda_{t_j}}^k}, \quad  i = 1, \dots, N
\end{equation}
Where $\Delta B^i_j$ are independent and identically distributed (i.i.d.) Normal Variables with $\mu = 0$ and $\sigma^2 = \Delta t$. Numerically, we have found this method to be effective as long as the starting particles start sufficiently far apart. We also found that the simulations worked with the points started close together, but often times the strong repulsion force lead to numerical issues. For our main results the points were successfully spread apart. We also chose a relatively short final time $T = 0.25$ which helped prevent numerical issues. As was already mentioned in the Introduction, this time was also motivated as it is known that for relatively short time scales on order of $\frac{1}{N}$ for Dyson Brownian Motion it is known to reach a local equilibrium (see \cite{DRMT}). 
It is known that the Loewner hull will not grow in height up to time $t \geq 0$, above $2i\sqrt{t}$ (see \cite{Gregconf}, \cite{Kemp}) and thus we will choose our starting points $z_0 \in \mathbb{H}$ outside of this area to ensure it is well defined. We will simulate our map by using another Euler Scheme using our previously sampled values. That is, if we let $g_t(z) = z_t$ we have
\begin{equation}
    z_{t_{j + 1}} = z_{t_j} + \frac{\Delta t}{N}\sum_{i=1}^{N}\frac{2}{z_{t_j} -\lambda_{t_j}^i}, \quad z_{t_0} = z_0
\end{equation}
An example pseudocode of our process is given in Algorithm \ref{alg:n} below. The source code can be obtained in the following repository\footnote{https://github.com/PHillytheBilly/SLEexploration}\\
\begin{algorithm}
    \caption{An algorithm to numerically simulate $g_t(z)$}\label{alg:n}
    \begin{algorithmic}
        \State Given $\Delta t$, $T$, $N$, $x_0$, and $z_0$
        \State $i \gets 1$
        \State $t = 0$
        \State $\lambda_t^i \gets x_0^i$ for $i = 1, \dots, N$
        \While{$t \leq T$}
            \State $t_n \gets t + \Delta t$
            \For{$i \leq N$}
                \State $\lambda_{t_n}^{i} = \lambda^i_t + \frac{\sqrt{2}}{\sqrt{N\beta}}\Delta B_j^i + \frac{1}{N} \sum_{j \neq k}^{N} \frac{\Delta t}{{\lambda_{t}}^j-{\lambda_{t}}^k}$
            \EndFor    
            \State $t \gets t_n$
        \EndWhile
        \State $t = 0$
        \While{$t \leq T$}
            \State $t_n \gets t + \Delta t$
            \State $z_{t_n} = z_{t} + \frac{\Delta t}{N}\sum_{j=1}^{N}\frac{2}{z_{t} -\lambda_{t}^j}, \quad z_{t_0} = z_0$
            \State $t \gets t_n$
        \EndWhile
        \State Return $z_T$
    \end{algorithmic}
\end{algorithm}
Another challenge with these dynamics when it comes to simulating them is the absence of closed-form solutions for fixed $N$ finite number of Dyson particles. However, there is close-form formula in the limit $N \to \infty.$ In order to test the accuracy of our algorithm, we use this fact and we showed that our simulations tend to approach the closed-form expression given by the theoretical solution, that is as $N$ increases our maps numerically converge to the map that is the limit of $N \to \infty$. In particular, it was shown in \cite{Hotta_2018} that as $N \to \infty$, we have
\begin{equation}
    g_t(z) = 2i\sqrt{t} \cdot \left(\frac{1}{W_0(-4t/z^2)} - W_0(-4t/z^2)\right)
\end{equation}

where $W_0$ is the Lambert $W$ function. Note that this only occurs when the DBM all starts from a single source. In our simulation, we start the DBM from very close to each-other and to the origin.  In this case, we found that there is very good alignment with the convergence towards the our theoretical map as seen in \ref{fig:EST} \\
In addition, we performed experiments in order to predict the distribution of $N|g_t(z) - \overline{g_t(z)}|$. In the previous formula, $\overline{g_t(z)}$ was calculated by $\overline{g_t(z)} = \frac{1}{\tilde{N}}\sum g_t(z)$ where $\tilde{N}$ is the number of experiments runned. As was already mentioned in the Introduction, this part was motivated by previous results which found that an analogous quantity for Dyson Brownian Motion converges to a Gaussian Process \cite{BENDER20081022}. To fit our distribution, we took the distribution with the lowest Sum Squared Error from a histogram after testing every distribution in SciPy's Library\cite{egCode}, with the exception of the Levy Stable and Studentized Range distributions. In particular, Levy Stable was excluded as it takes much longer to fit than other distributions. The results of these experiments and the corresponding predictions are in the Appendix.
\begin{figure}[H]
    \centering
    \includegraphics[scale=0.4]{
        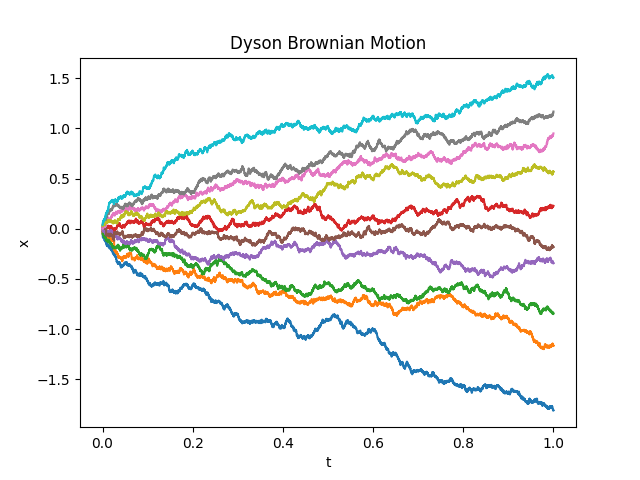
        }
    \caption{An example of Dyson Brownian Motion for $\beta=2$ for $N=10$ with all points starting roughly at $0$.}
    \label{fig:enter-label}
\end{figure}
\section{Numerical simulations for the SLE maps dynamics}
As we are interested in the behaviour of our quantities as the parameters are varied we ran experiments for $\kappa = 1, 2, \frac{8}{3}, 3, 4, 5$ with $N = 1$. We also looked at the trajectories of these maps starting from $z_0 = 1.02i$ and $z_0 = 3i$. These points were chosen for reference to try and see behaviors near the theoretical hull and further from this hull. As it was explained in the Introduction, other points could have been chosen, and we may look at other points in the future. To do this, we generated 5000 instances of Brownian Motion starting from $x = 0$, and then used these samplings to run our $g_{0.25}(z_0)$. \\
\subsection{The SLE maps dynamics stared from $z_0 = 1.02i$}
When starting the maps close to the $2i\sqrt{t}$ boundary, our simulations predicted that the final trajectories would land in a bimodal distribution, most often described by a Wrapped Cauchy distribution. An example image of this distribution is shown in \ref{fig:SLE1} and more examples are given in Appendix \ref{app:SLE1}. 
\begin{figure}[H]
        \centering
        \includegraphics[width=\textwidth]{
            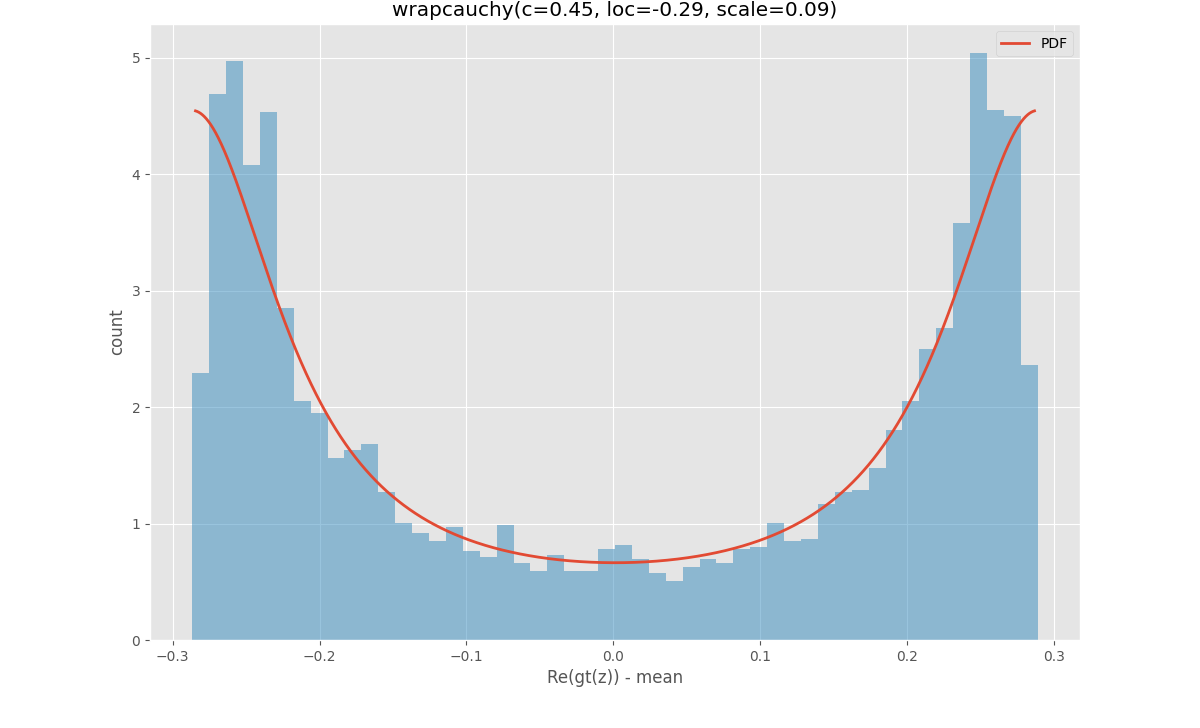
        }
    \caption{Histogram of $Re(g_{0.25}(1.02i)) - \overline{Re(g_{0.25}(1.02i))}$ for $\kappa = \frac{8}{3}$.}
    \label{fig:SLE1}
\end{figure}
\subsection{The SLE maps dynamics started from $z_0 = 3i$}
When starting our maps further up, we found that the values were much more closely distributed around the sample mean, which suggests a tighter clustering. This is represented by a "bell shape" in $Re(g_t(z)) - \overline{Re(g_t(z)}$. This matches the intuition since as $|z| \to \infty$, $g_t(z) - z \to 0$ \cite{notes} which suggests that the further we move away from the origin, the "nicer" our map behaves (see \cite{Gregconf}, \cite{Kemp} for more details). The distributions for other values of $\beta$ can be found in Appendix \ref{app:SLE3}.
\begin{figure}[H]
    \centering
       \includegraphics[width=\textwidth]{
           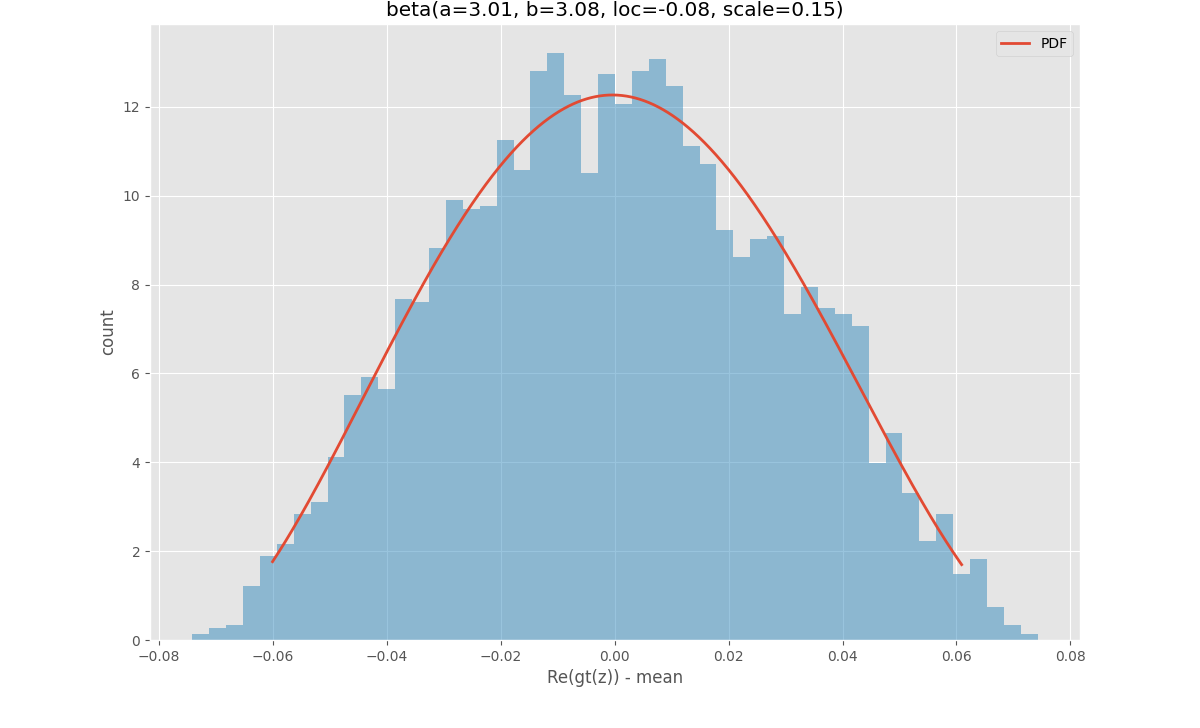
       }
    \caption{Histogram of $Re(g_{0.25}(3i)) - \overline{Re(g_{0.25}(1.02i))}$ for $\kappa = \frac{8}{3}$.}
    \label{fig:SLE3}
\end{figure}
\section{Numerical simulations for the Multiple SLE dynamics}
\begin{figure}[H]
    \centering
    \includegraphics[width=0.5\textwidth]{
        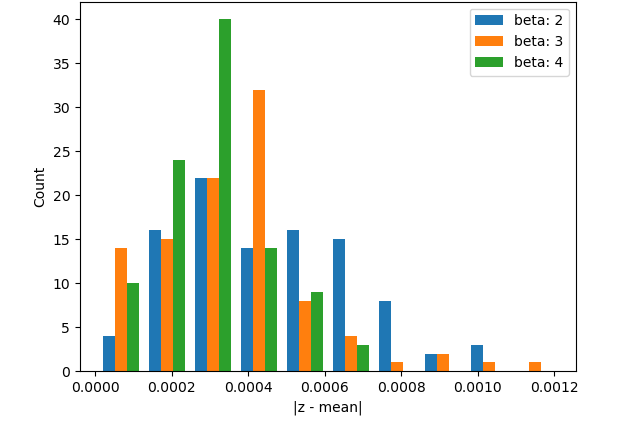
    }
    \caption{Histogram of $|g_t(z) - \overline{g_t(z)}|$ for $\beta = 2, 3, 4$.}
    \label{fig:bdepend}
\end{figure}

In the case of the Multiple SLE dynamics the computations cost is much higher given the complexity of both the DBM drivers dynamics and their impact on the multiple SLE. The first thing to note was that for $N=60$, that there was a dependence on $\beta$ for how spread the values were, as seen in \ref{fig:bdepend}. This analysis is done with $100$ experiments with dynamics starting from $z_0 = 1 + 2i$. \\
These experiments show a good prediction with the theory as when $\beta$ decreases the parameter $\kappa$ increases (as $\beta=8/\kappa$) and we see this reflected in the simulations as well according to the image. There is more spread around the sample mean as $\kappa$ is larger. 
We ran further experiments with $\beta = 1, 2$ for $N = 10, 50, 100$. To select the initial $X_0$ for our Dyson Brownian Motion, we had all points start around $0$ with a distance of $0.2$ between each point. This was done to ensure that sufficient space was between the starting points so that numerical issues were less likely from the repulsion force. We also looked at the starting points $z_0 = 1.02i, 3i$. For each experiment we ran 500 instances. All of the experiments predicted a bell shaped curve for the real parts, and a skewed left distribution for the absolute values. We hope that these predictions will initiate more theoretical work to investigate further these distributions. An example for $\beta=1$ and $N=50$ is shown in \ref{fig:DBM1}. Additional figures of the distributions are provided in Appendix \ref{app:DBM1} and Appendix \ref{app:DBM3}.
\begin{figure}[H]
    \begin{subfigure}{0.5\textwidth}
        \centering
        \includegraphics[width=\textwidth]{
            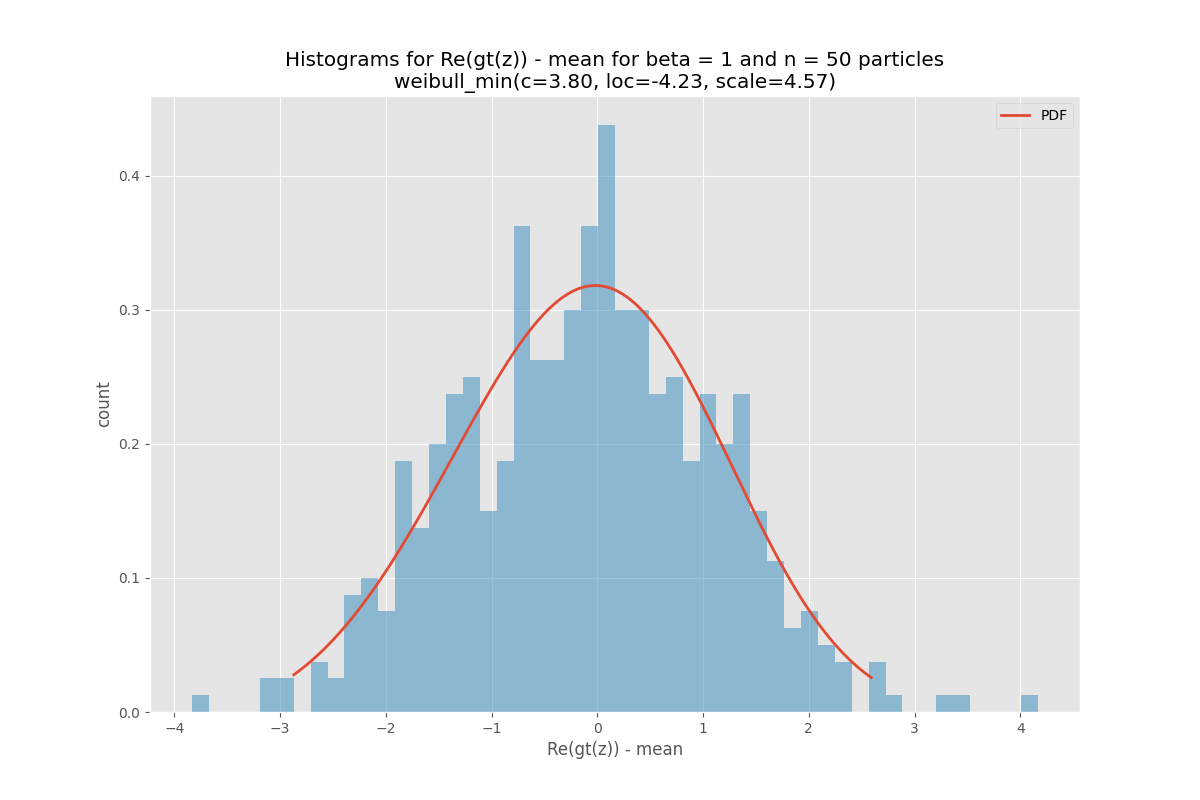
        }
        \caption{$N \cdot(Re(g_{0.25}(1.02i)) - \overline{Re(g_{0.25}(1.02i))})$}
    \end{subfigure}
    \begin{subfigure}{0.5\textwidth}
        \centering
       \includegraphics[width=\textwidth]{
           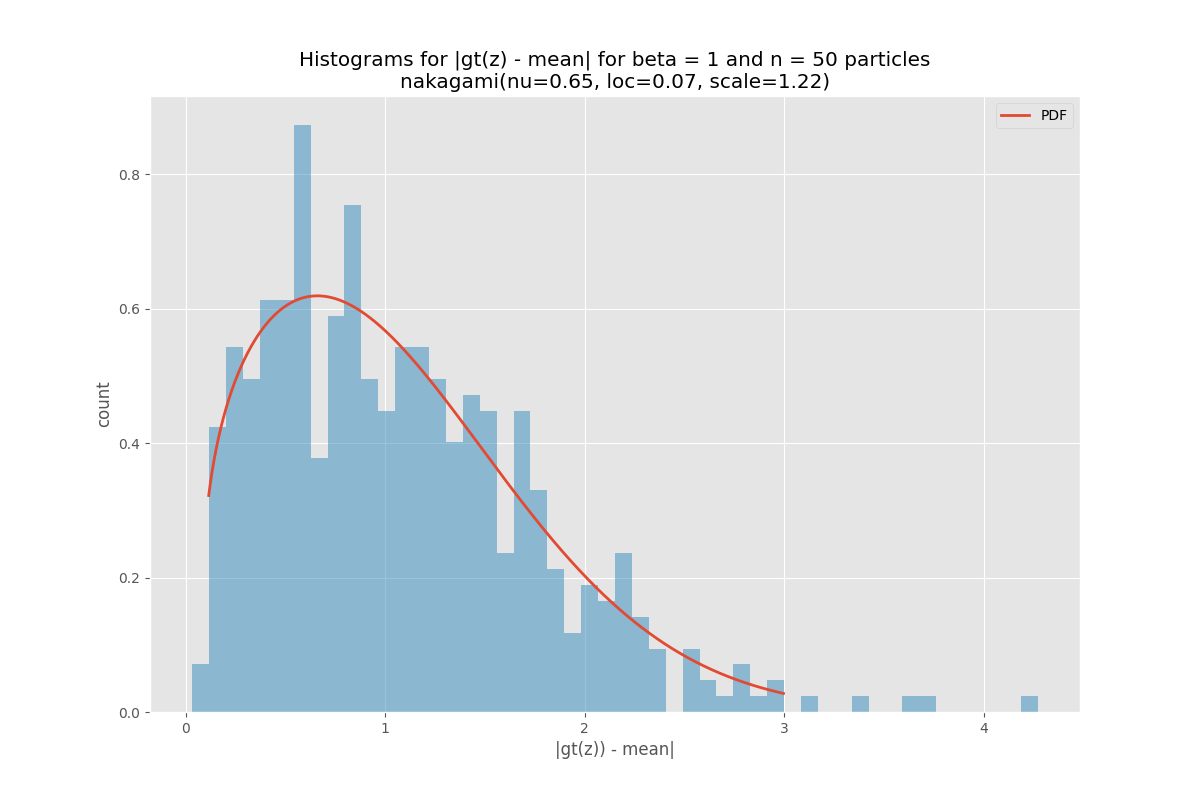
       }
       \caption{$N \cdot |g_{0.25}(1.02i) - \overline{g_{0.25}(1.02i)}|$}
    \end{subfigure}
    \caption{Histograms for $\beta = 1$ and $N = 50$.}
    \label{fig:DBM1}
\end{figure}

\section{Future Directions}

We hope that these explorations will stimulate theoretical work in the future to check these numerical predictions. In addition, we encourage the study of the spread around the average not only at a fixed time but also as a process of time. In addition, we believe it is beneficial to expand the analysis to include the imaginary parts rather than looking just at the real parts. This will consist of studying the spread of the dynamics from the average in the Complex Plane and its dependency on the parameter $\kappa \geq 0$ as processes in time. Our experiments show that for a fixed time the spread of the points is larger as $\beta$ decreases (that is, as $\kappa$ increases) as it is expected given that the noise is amplified as the $\kappa$ increases. 
\begin{figure}[H]
    \centering
    \includegraphics[width=0.5\linewidth]{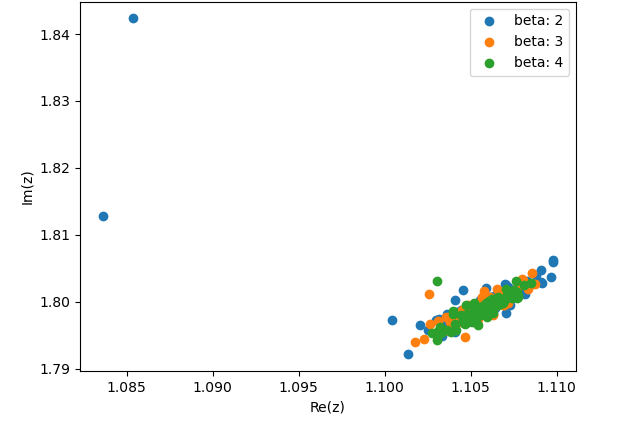}
    \caption{Example plot within the Complex Plane with starting point with non-zero real part. }
    \label{fig:cloud}
\end{figure}

\section{Acknowledgements}V.M. would like to thank A. Campbell, K. Luh and T. Alberts for useful discussions. 
P.K and V.M. also acknowledge that this work was funded by the NSF-REU DMS-2150179 grant.
\printbibliography
\appendix
\section[Appendix A]{Numerical experiments studying the spread for the SLE maps dynamics started from $z_0 = 1.02i$} \label{app:SLE1}
\begin{figure}[H]
    \begin{subfigure}{0.5\textwidth}
        \centering
        \includegraphics[width=\textwidth]{
            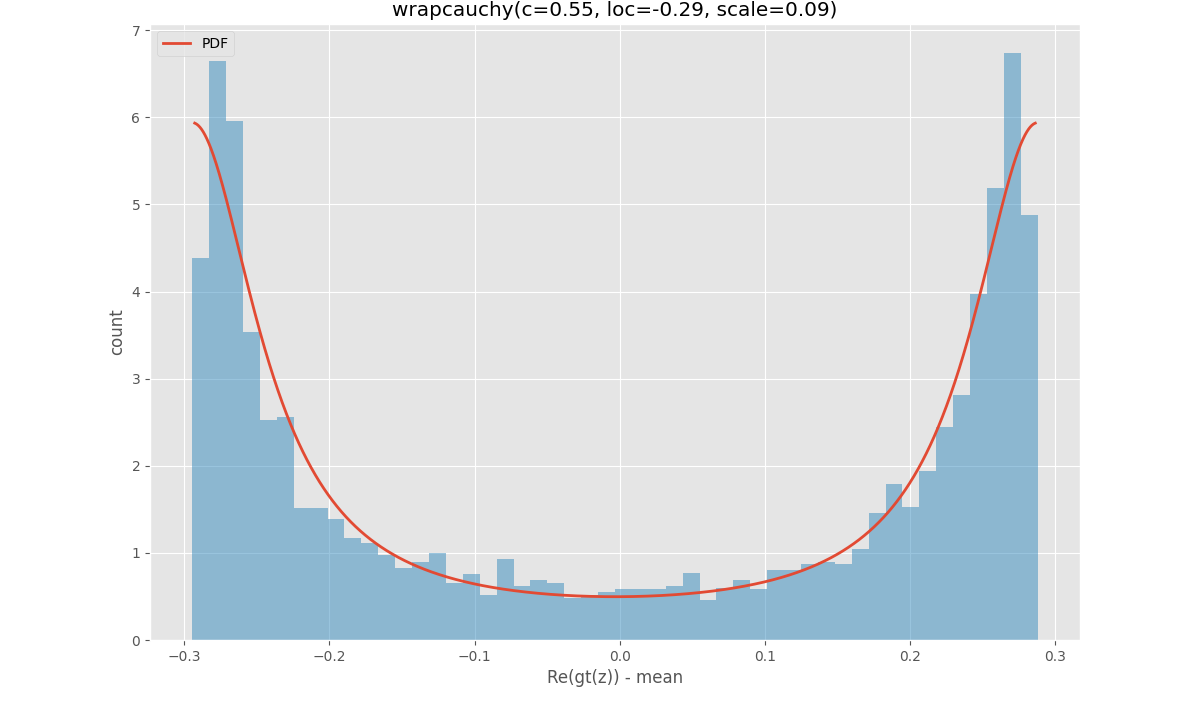
        }
        \caption{$\kappa = 1$}
        \label{fig:SLE11}
    \end{subfigure}
    \begin{subfigure}{0.5\textwidth}
        \centering
       \includegraphics[width=\textwidth]{
           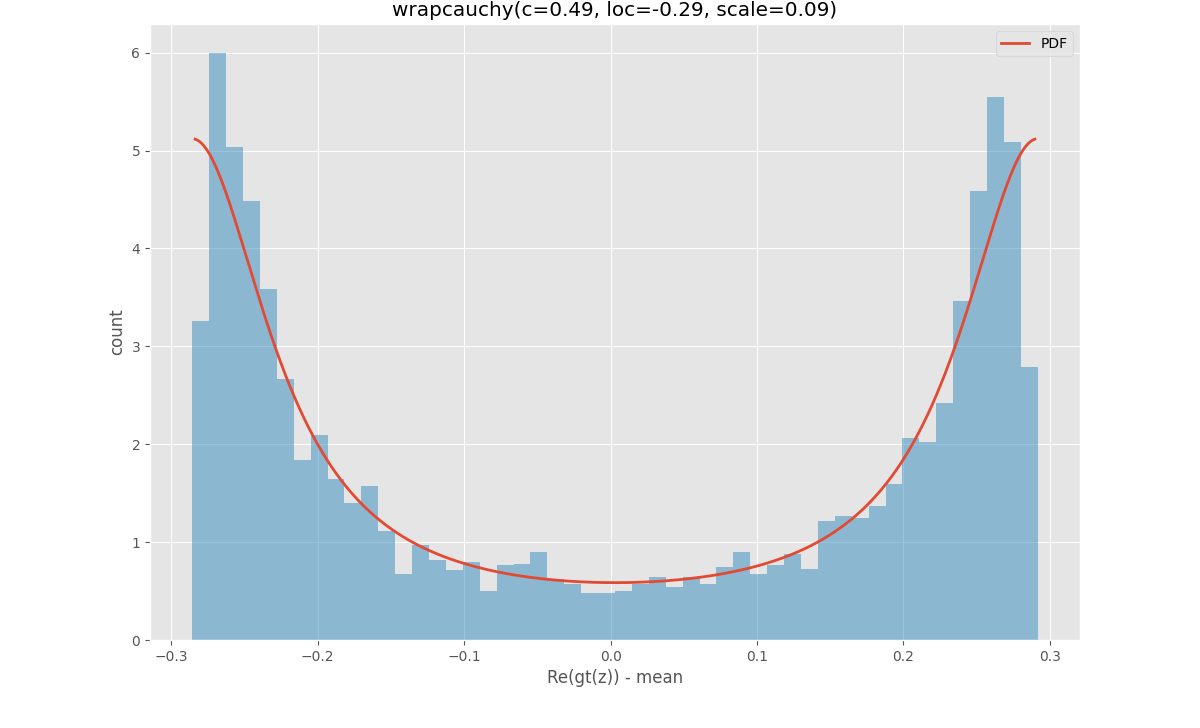
       }
       \caption{$\kappa = 2$.}
       \label{fig:SLE21}
    \end{subfigure}
    \caption{Histograms for $Re(g_{0.25}(1.02i)) - \overline{Re(g_{0.25}(1.02i))}$.}
\end{figure}
\begin{figure}[H]
    \begin{subfigure}{0.5\textwidth}
        \centering
        \includegraphics[width=\textwidth]{
            imagesRealLargeC/3/R_1_bestdist_cropped.png
        }
        \caption{$\kappa = \frac{8}{3}$}
        \label{fig:SLE31}
    \end{subfigure}
    \begin{subfigure}{0.5\textwidth}
        \centering
       \includegraphics[width=\textwidth]{
           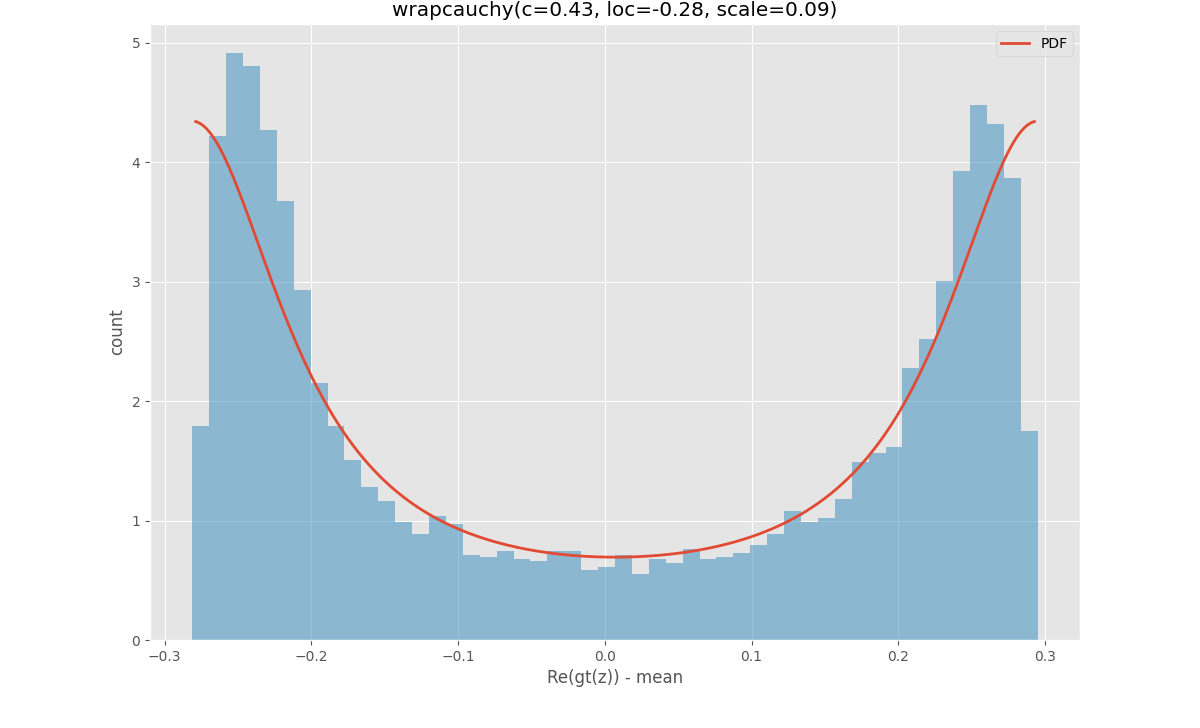
       }
       \caption{$\kappa = 3$.}
       \label{fig:SLE41}
    \end{subfigure}
    \caption{Histograms for $Re(g_{0.25}(1.02i)) - \overline{Re(g_{0.25}(1.02i))}$.}
\end{figure}
\begin{figure}[H]
    \begin{subfigure}{0.5\textwidth}
        \centering
        \includegraphics[width=\textwidth]{
            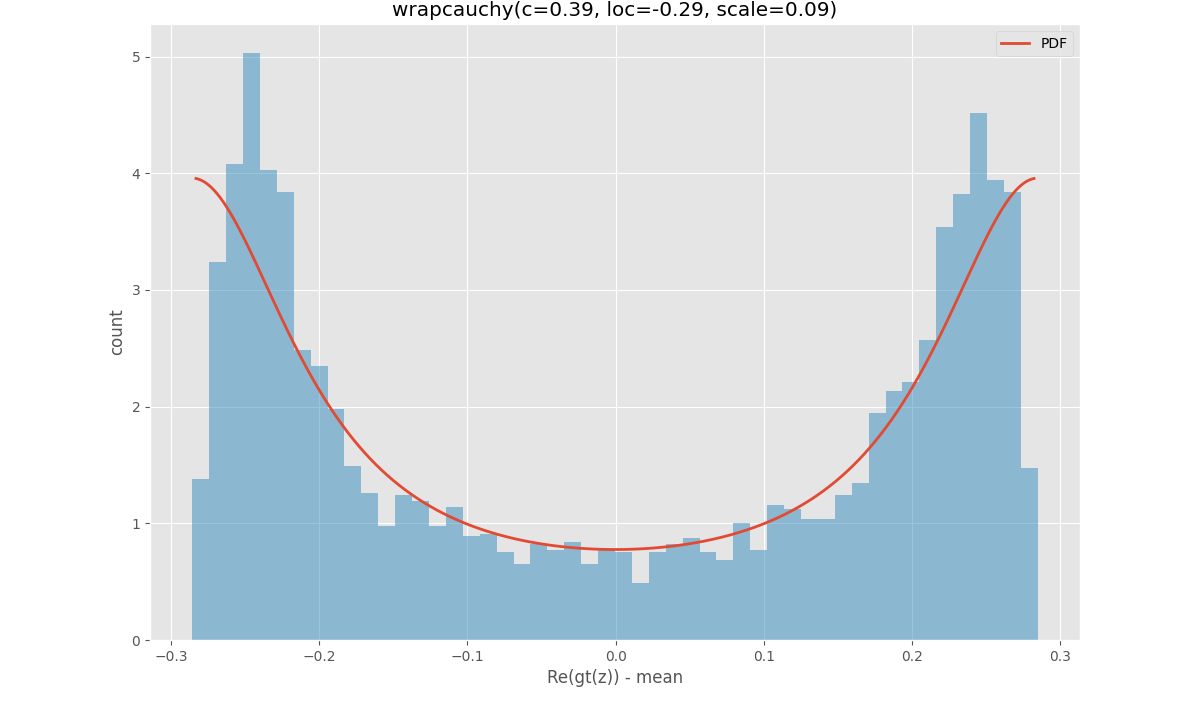
        }
        \caption{$\kappa = 4$}
        \label{fig:SLE51}
    \end{subfigure}
    \begin{subfigure}{0.5\textwidth}
        \centering
       \includegraphics[width=\textwidth]{
            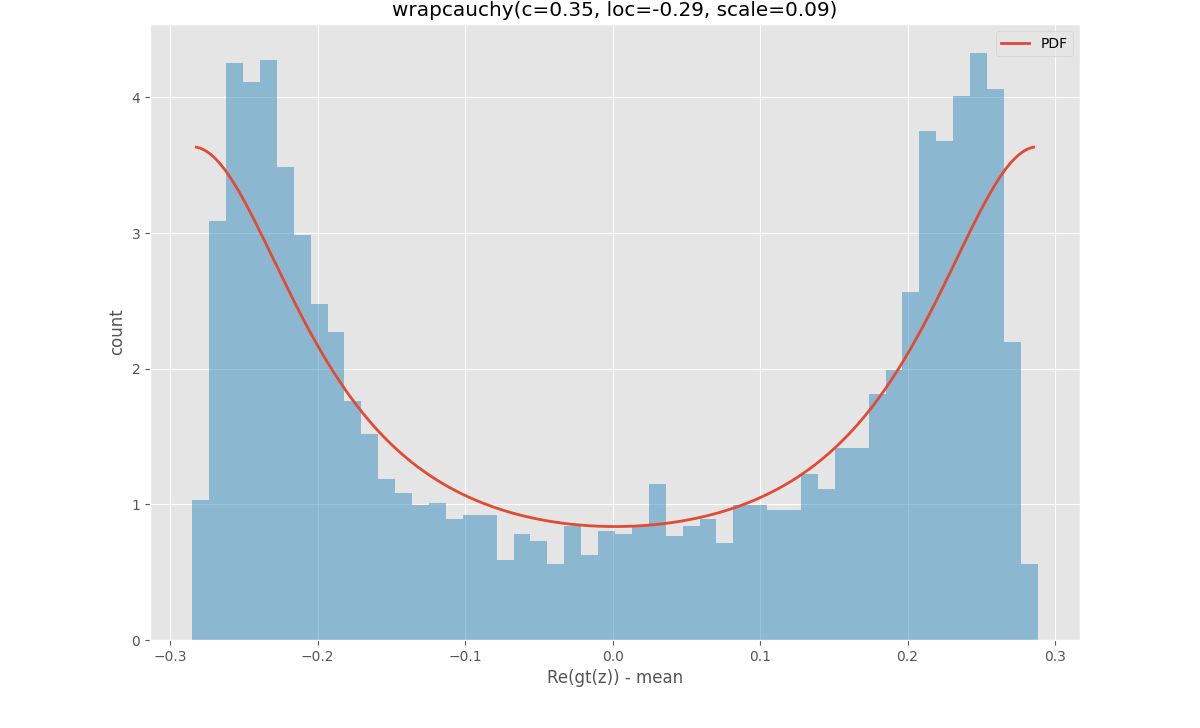
       }
       \caption{$\kappa = 5$.}
       \label{fig:SLE61}
    \end{subfigure}
    \caption{Histograms for $Re(g_{0.25}(1.02i)) - \overline{Re(g_{0.25}(1.02i))}$.}
\end{figure}

\section{Numerical experiments studying the spread of the dynamics for the SLE maps dynamics started from $z_0 = 3i$} \label{app:SLE3}

\begin{figure}[H]
    \begin{subfigure}{0.5\textwidth}
        \centering
        \includegraphics[width=\textwidth]{
            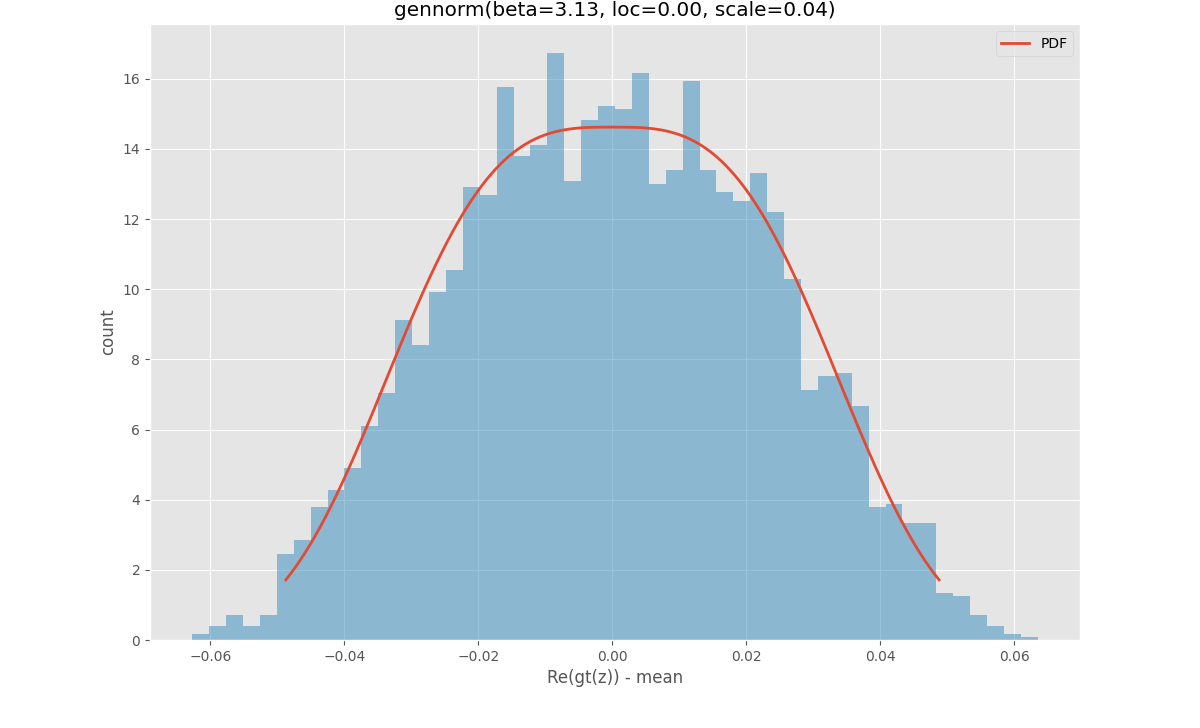
        }
        \caption{$\kappa = 1$}
        \label{fig:SLE12}
    \end{subfigure}
    \begin{subfigure}{0.5\textwidth}
        \centering
       \includegraphics[width=\textwidth]{
            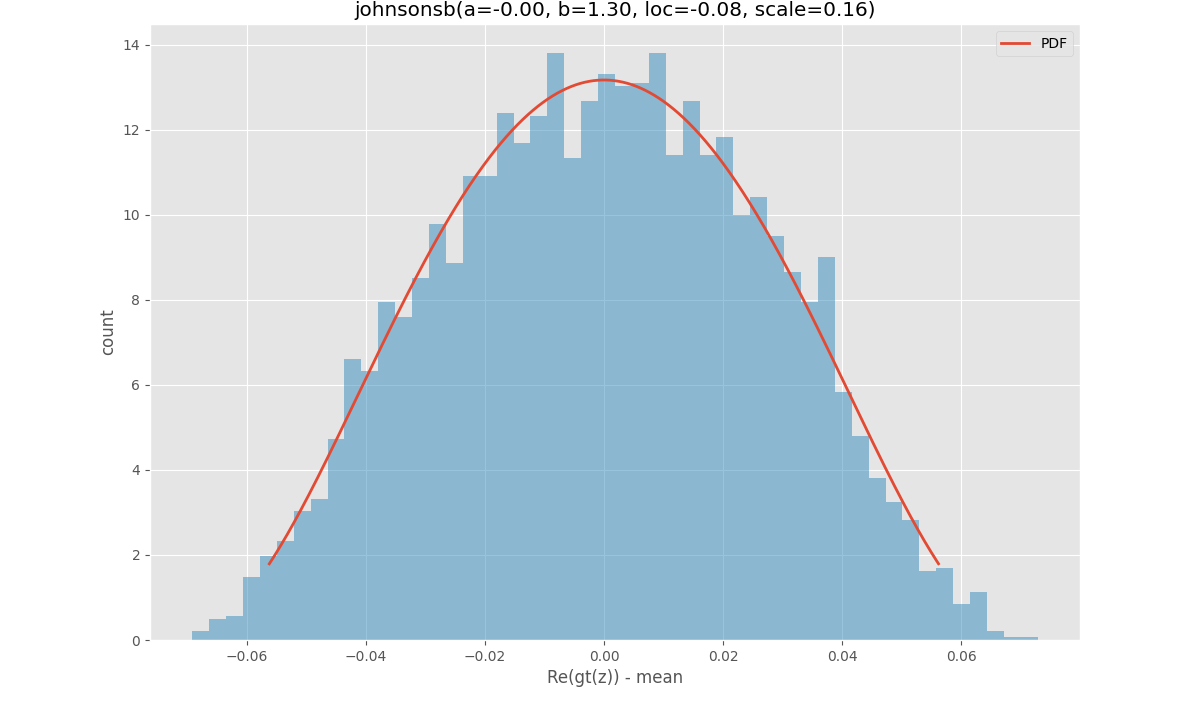
       }
       \caption{$\kappa = 2$.}
       \label{fig:SLE22}
    \end{subfigure}
    \caption{Histogram of $Re(g_{0.25}(3i)) - \overline{Re(g_{0.25}(3i))}$.}
\end{figure}
\begin{figure}[H]
    \begin{subfigure}{0.5\textwidth}
        \centering
        \includegraphics[width=\textwidth]{
            imagesRealLargeC/3_2/R_1_bestdist_cropped.png
        }
        \caption{$\kappa = \frac{8}{3}$}
        \label{fig:SLE32}
    \end{subfigure}
    \begin{subfigure}{0.5\textwidth}
        \centering
       \includegraphics[width=\textwidth]{
            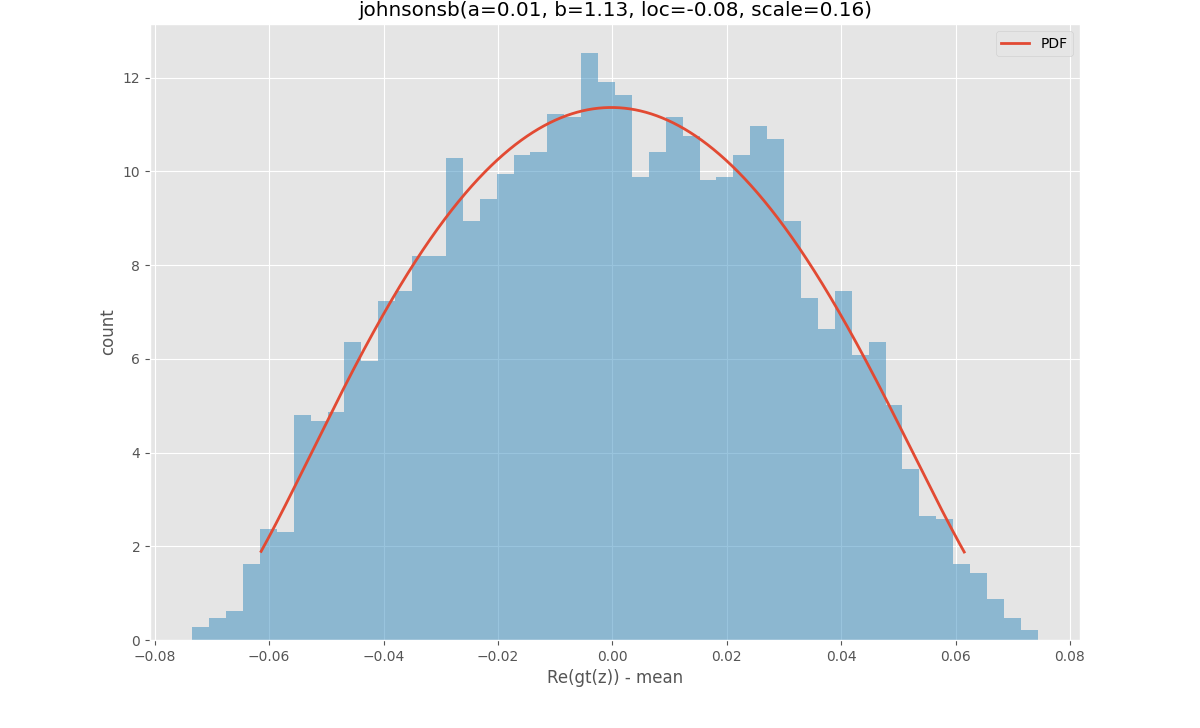
       }
       \caption{$\kappa = 3$.}
       \label{fig:SLE42}
    \end{subfigure}
    \caption{Histogram of $Re(g_{0.25}(3i)) - \overline{Re(g_{0.25}(3i))}$.}
\end{figure}
\begin{figure}[H]
    \begin{subfigure}{0.5\textwidth}
        \centering
        \includegraphics[width=\textwidth]{
            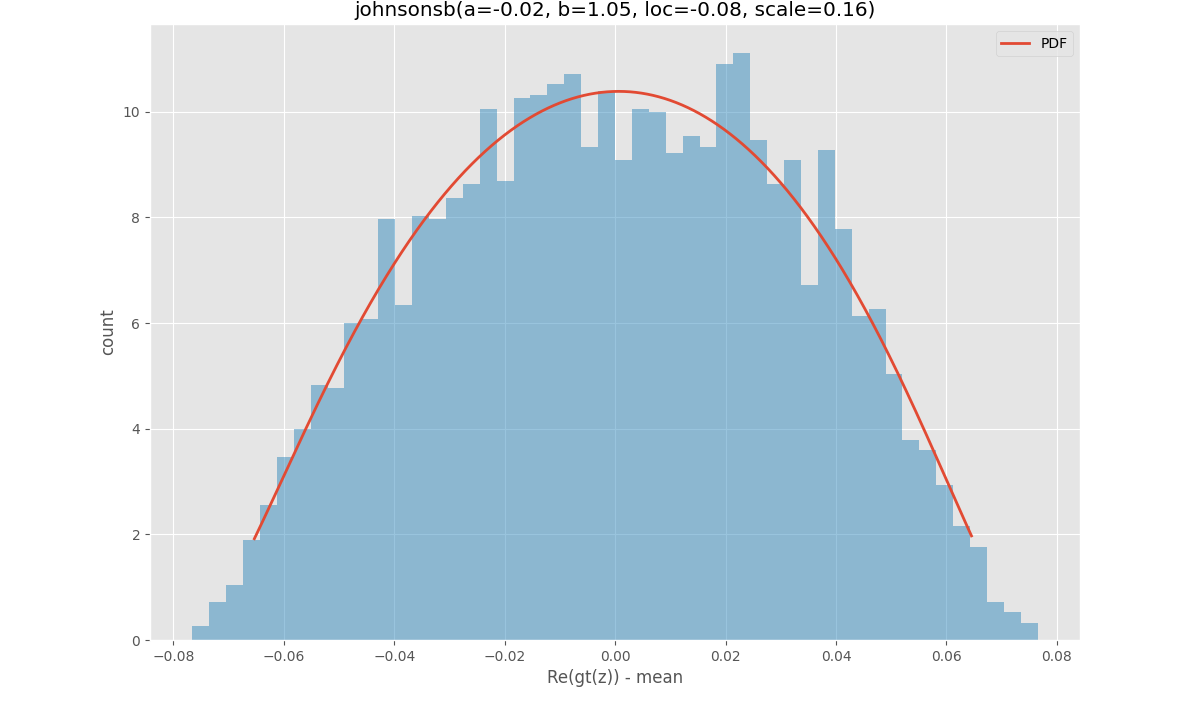
        }
        \caption{$\kappa = 4$}
        \label{fig:SLE52}
    \end{subfigure}
    \begin{subfigure}{0.5\textwidth}
        \centering
       \includegraphics[width=\textwidth]{
            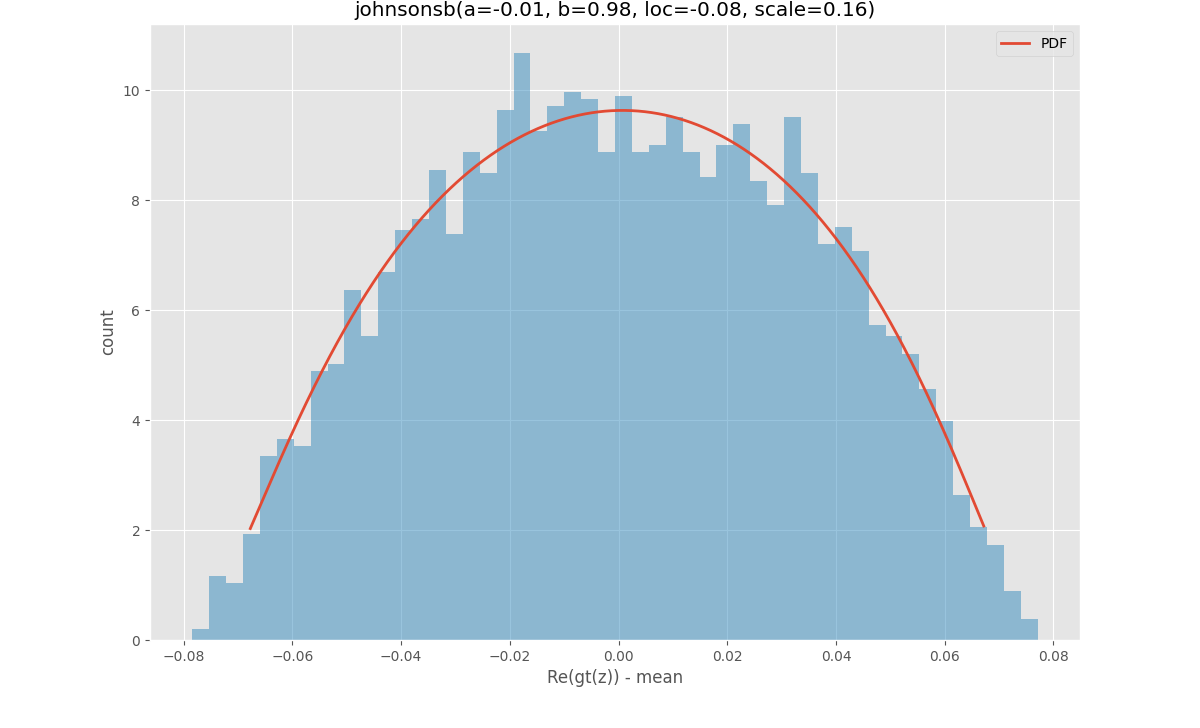
       }
       \caption{$\kappa = 5$.}
       \label{fig:SLE62}
    \end{subfigure}
    \caption{Histogram of $Re(g_{0.25}(3i)) - \overline{Re(g_{0.25}(3i))}$.}
\end{figure}

\section{Numerical experiments involving the spread for the mulitple SLE maps dynamics started from $z_0 = 1.02i$} \label{app:DBM1}
\begin{figure}[H]
    \begin{subfigure}{0.5\textwidth}
        \centering
        \includegraphics[width=\textwidth]{
            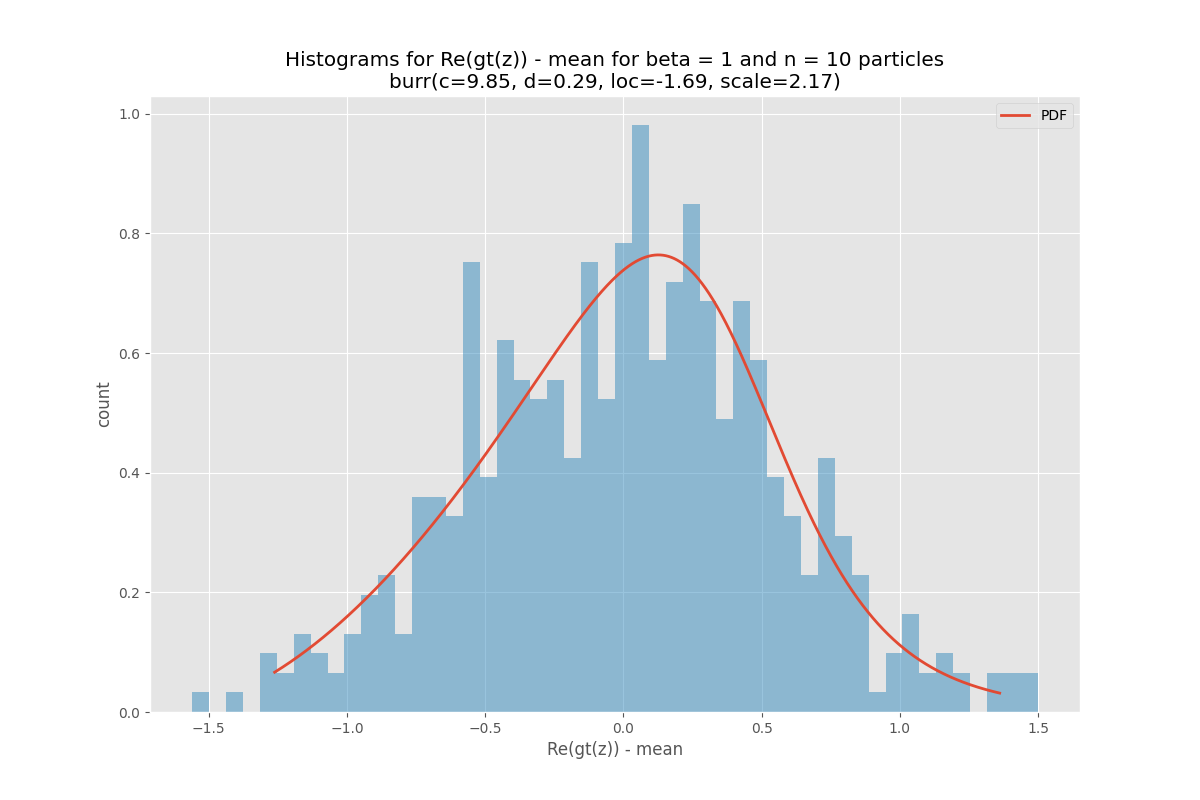
        }
        \caption{$N \cdot(Re(g_{0.25}(1.02i)) - \overline{Re(g_{0.25}(1.02i))})$}
    \end{subfigure}
    \begin{subfigure}{0.5\textwidth}
        \centering
       \includegraphics[width=\textwidth]{
           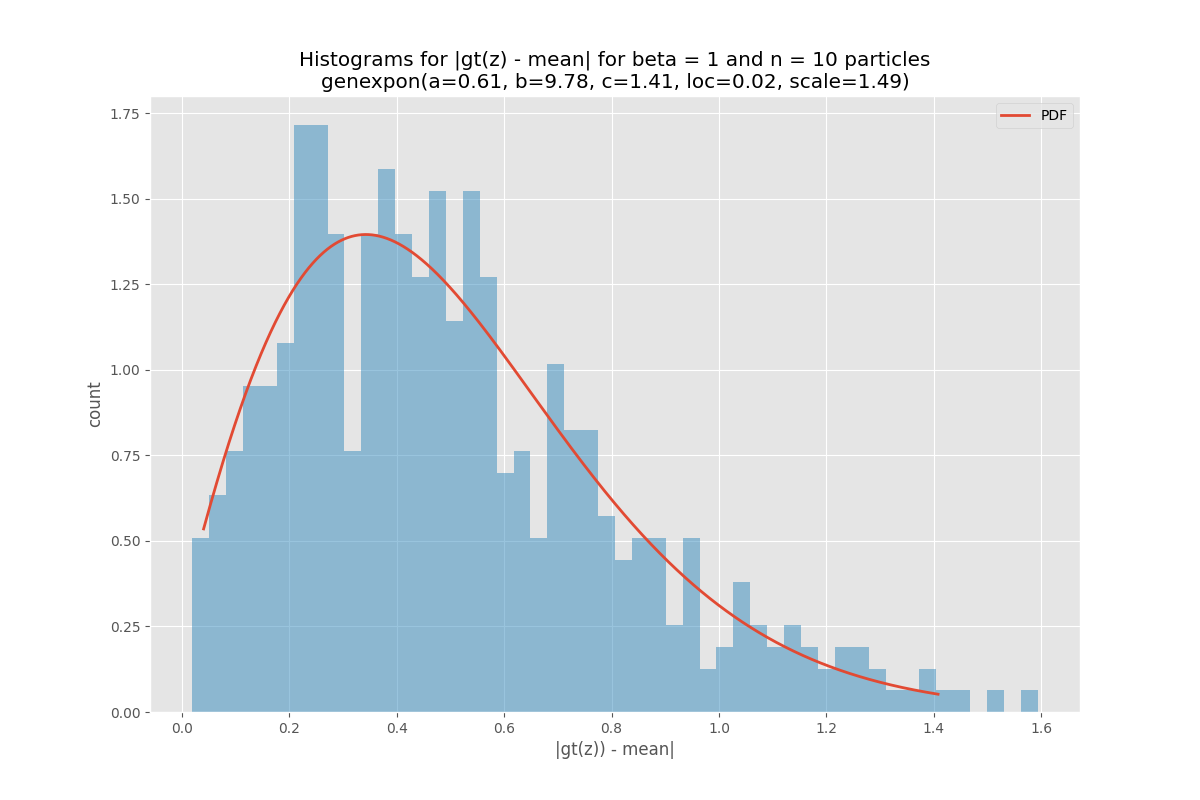
       }
       \caption{$N \cdot |g_{0.25}(1.02i) - \overline{g_{0.25}(1.02i)}|$}
    \end{subfigure}
    \caption{Histograms for $\beta = 1$ and $N = 10$.}
    \label{fig:1DBM1_10}
\end{figure}
\begin{figure}[H]
    \begin{subfigure}{0.5\textwidth}
        \centering
        \includegraphics[width=\textwidth]{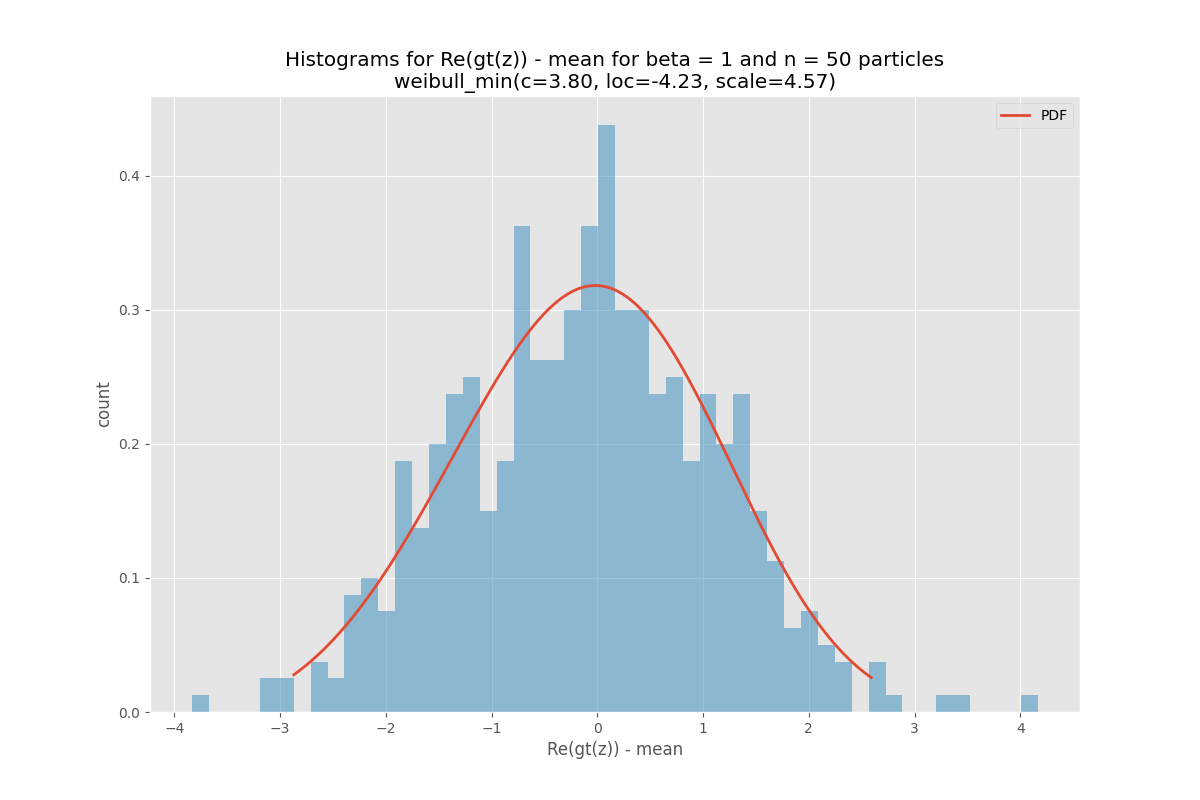}
        \caption{$N \cdot(Re(g_{0.25}(1.02i)) - \overline{Re(g_{0.25}(1.02i))})$}
    \end{subfigure}
    \begin{subfigure}{0.5\textwidth}
        \centering
       \includegraphics[width=\textwidth]{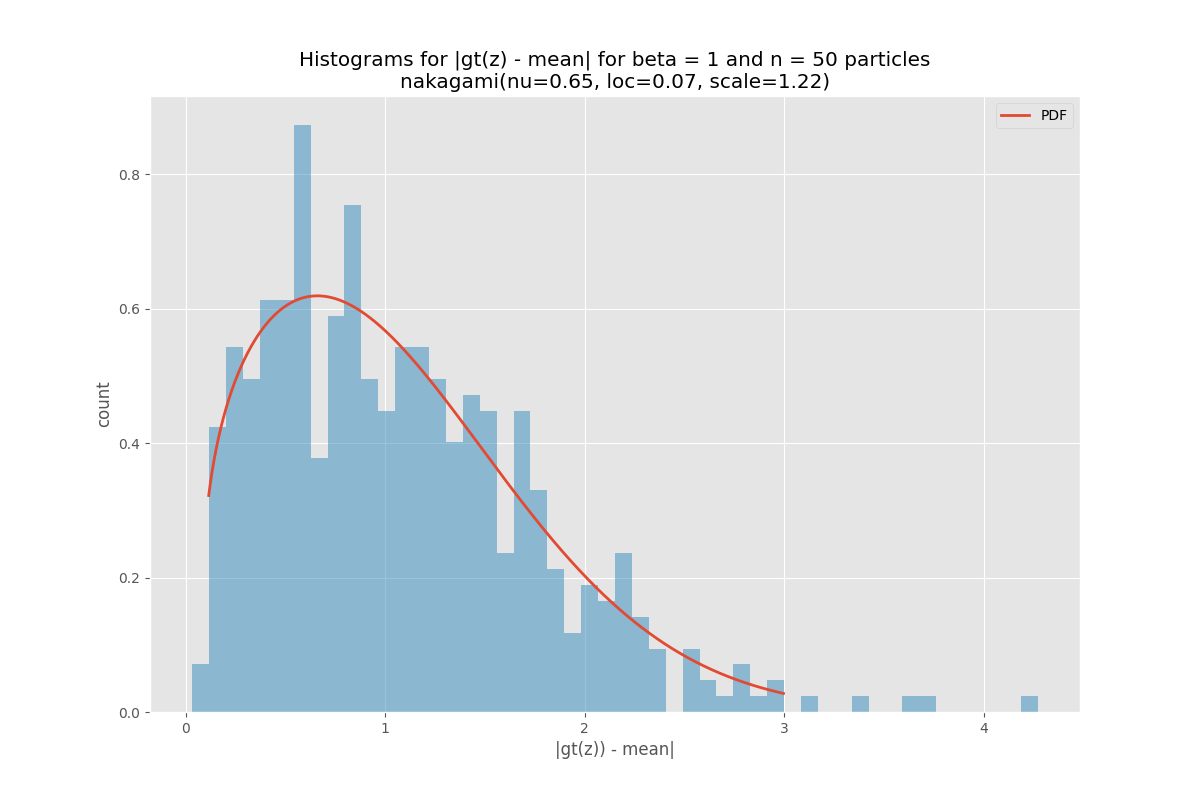}
       \caption{$N \cdot |g_{0.25}(1.02i) - \overline{g_{0.25}(1.02i)}|$}
    \end{subfigure}
    \caption{Histograms for $\beta = 1$ and $N = 50$.}
    \label{fig:1DBM1_50}
\end{figure}
\begin{figure}[H]
    \begin{subfigure}{0.5\textwidth}
        \centering
        \includegraphics[width=\textwidth]{
            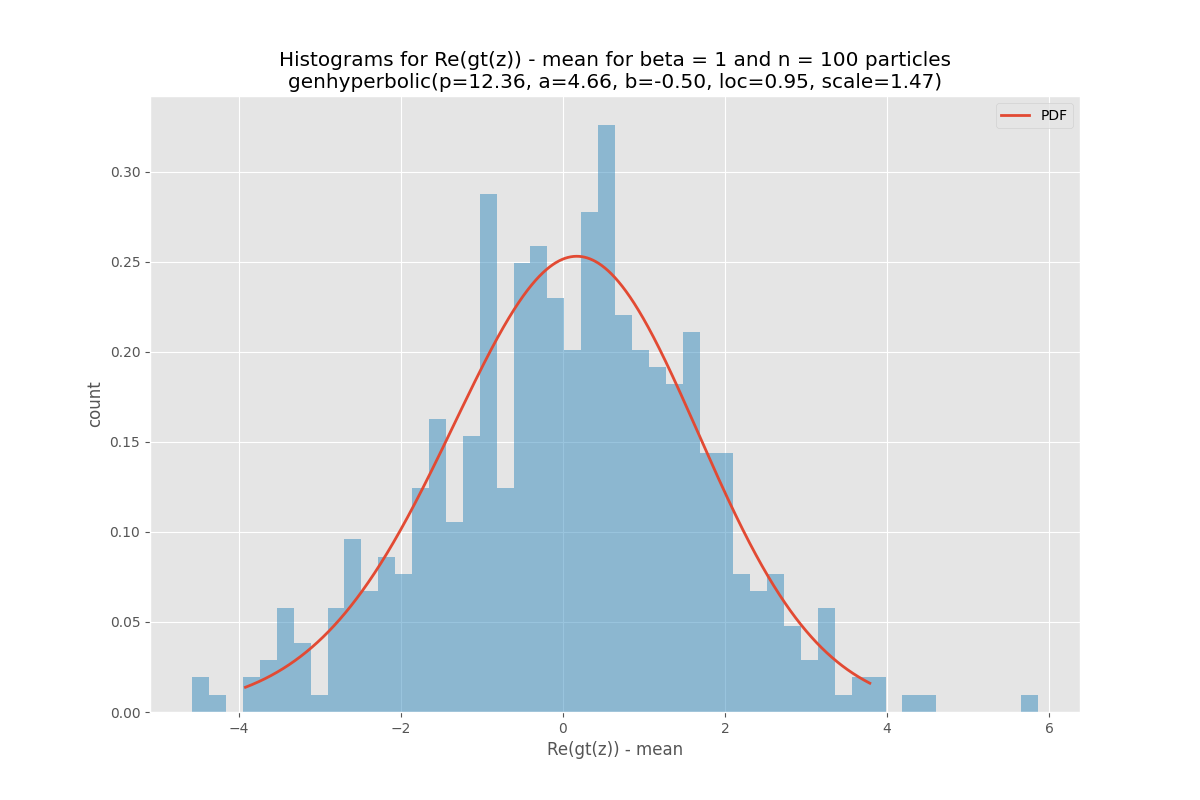
        }
        \caption{$N \cdot(Re(g_{0.25}(1.02i)) - \overline{Re(g_{0.25}(1.02i))})$}
    \end{subfigure}
    \begin{subfigure}{0.5\textwidth}
        \centering
       \includegraphics[width=\textwidth]{
           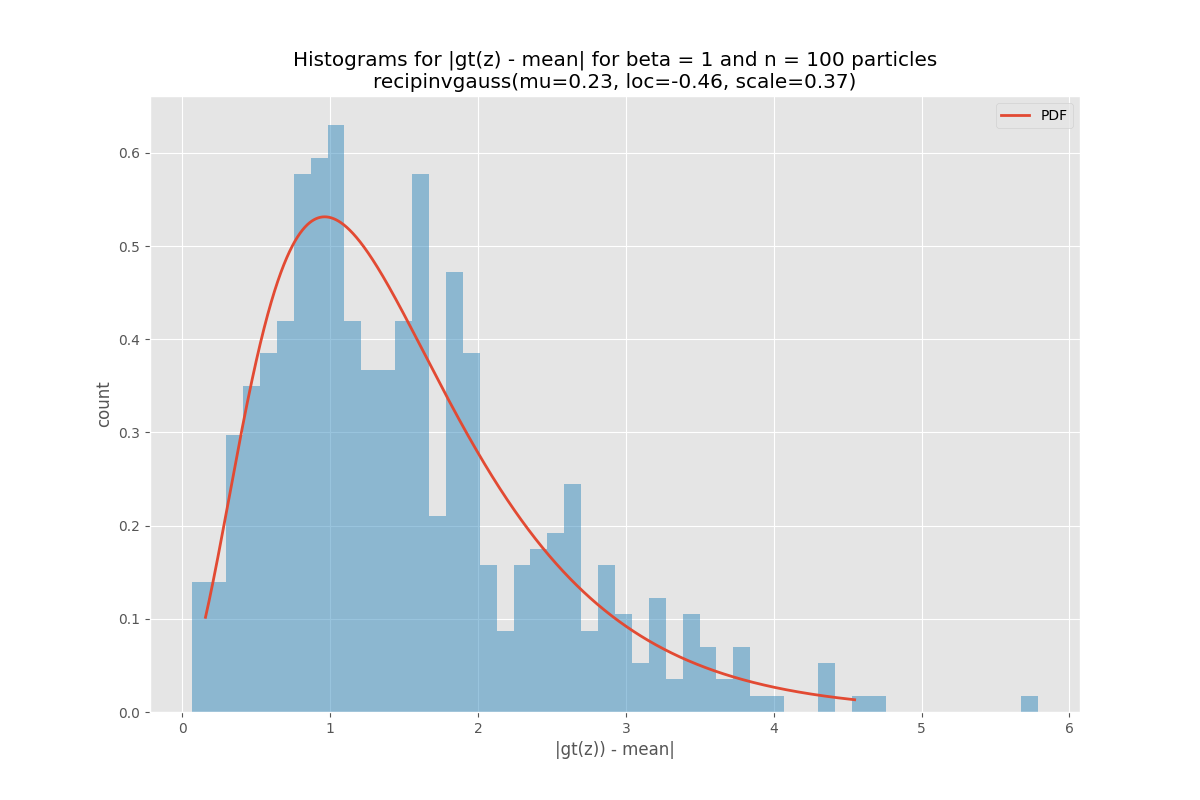
       }
       \caption{$N \cdot |g_{0.25}(1.02i) - \overline{g_{0.25}(1.02i)}|$}
    \end{subfigure}
    \caption{Histograms for $\beta = 1$ and $N = 100$.}
    \label{fig:1DBM1_100}
\end{figure}
\begin{figure}[H]
    \begin{subfigure}{0.5\textwidth}
        \centering
        \includegraphics[width=\textwidth]{
            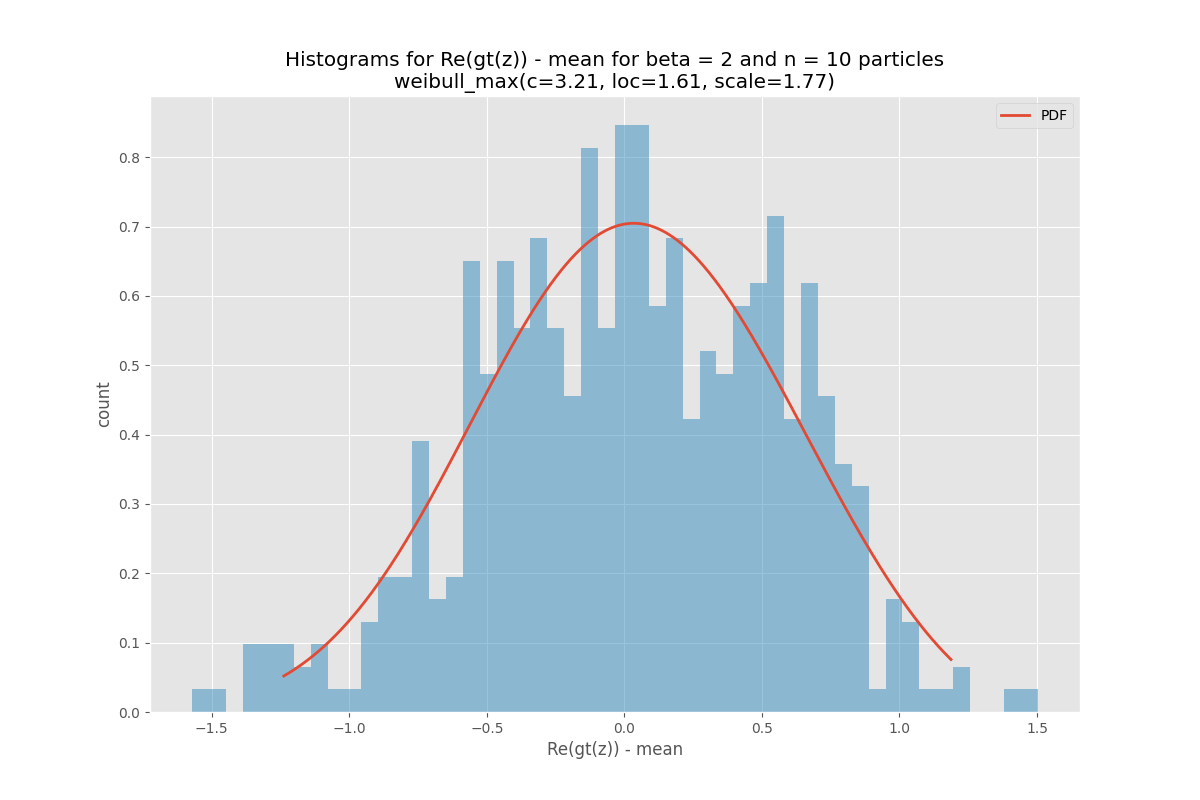
        }
        \caption{$N \cdot(Re(g_{0.25}(1.02i)) - \overline{Re(g_{0.25}(1.02i))})$}
    \end{subfigure}
    \begin{subfigure}{0.5\textwidth}
        \centering
       \includegraphics[width=\textwidth]{
           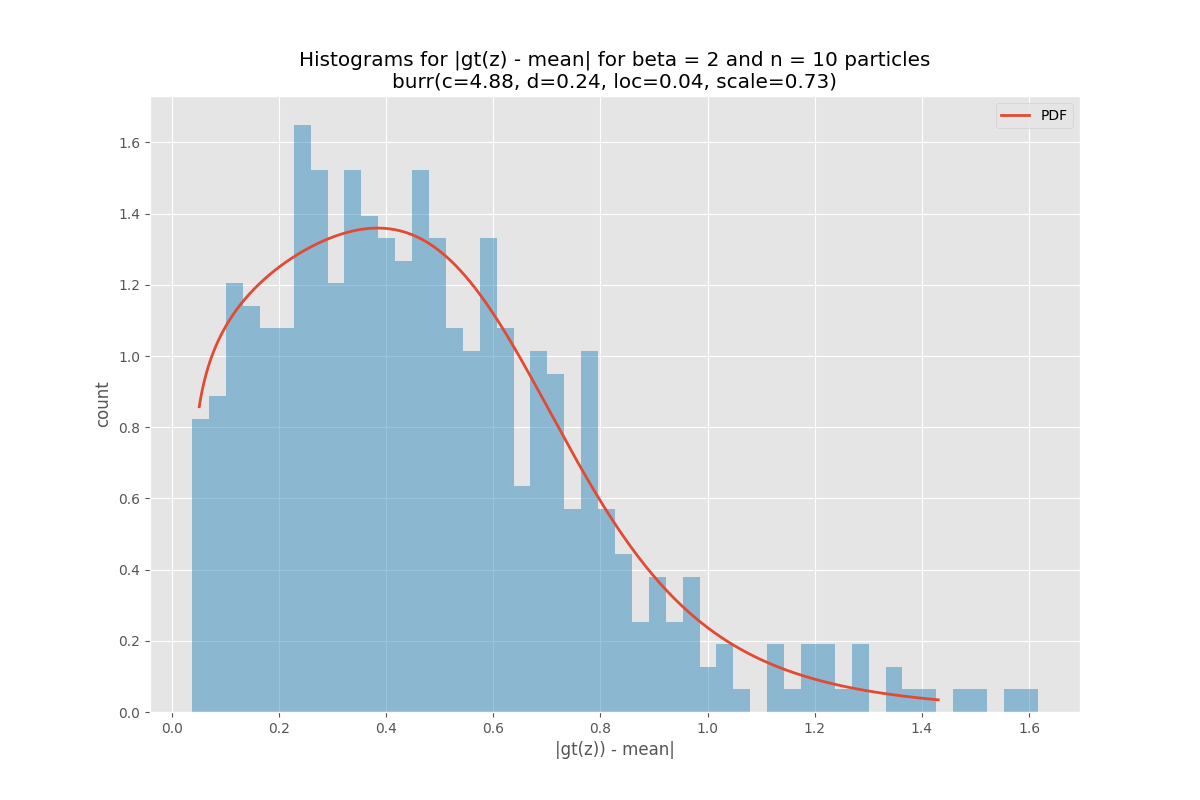
       }
       \caption{$N \cdot |g_{0.25}(1.02i) - \overline{g_{0.25}(1.02i)}|$}
    \end{subfigure}
    \caption{Histograms for $\beta = 2$ and $N = 10$.}
    \label{fig:1DBM2_10}
\end{figure}
\begin{figure}[H]
    \begin{subfigure}{0.5\textwidth}
        \centering
        \includegraphics[width=\textwidth]{
            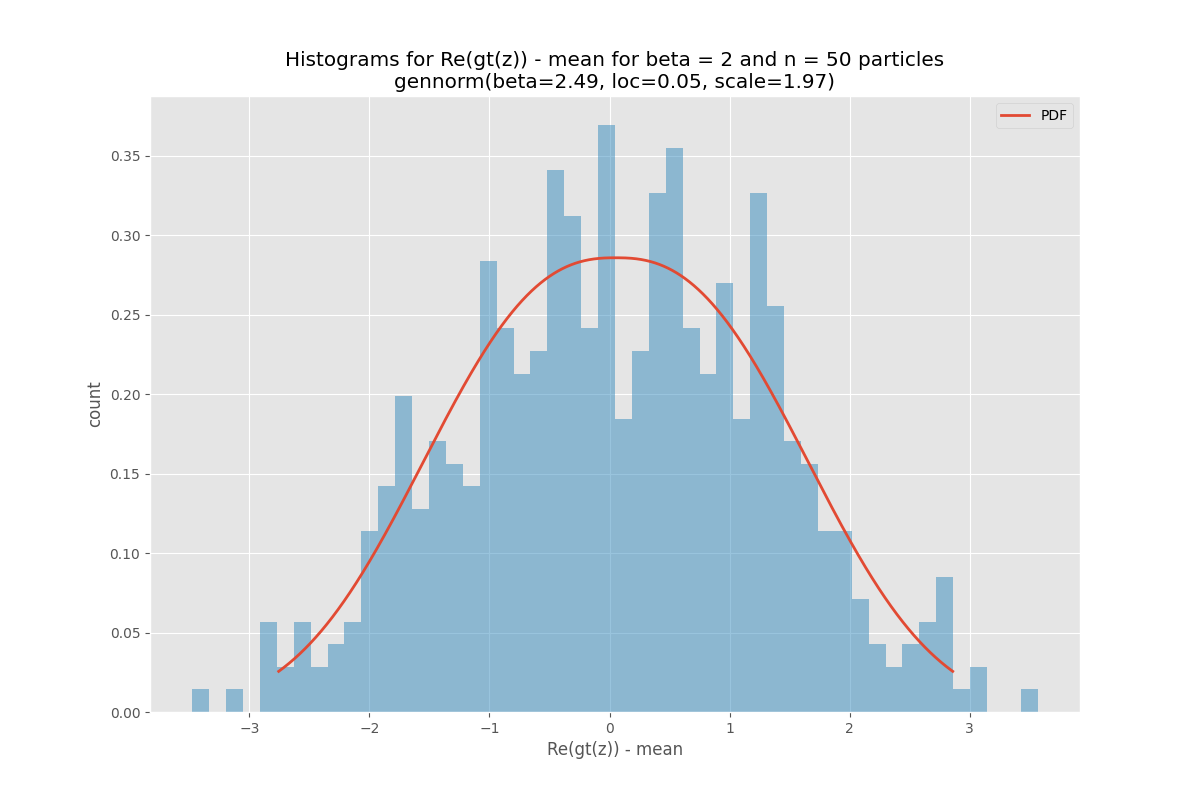
        }
        \caption{$N \cdot(Re(g_{0.25}(1.02i)) - \overline{Re(g_{0.25}(1.02i))})$}
    \end{subfigure}
    \begin{subfigure}{0.5\textwidth}
        \centering
       \includegraphics[width=\textwidth]{
           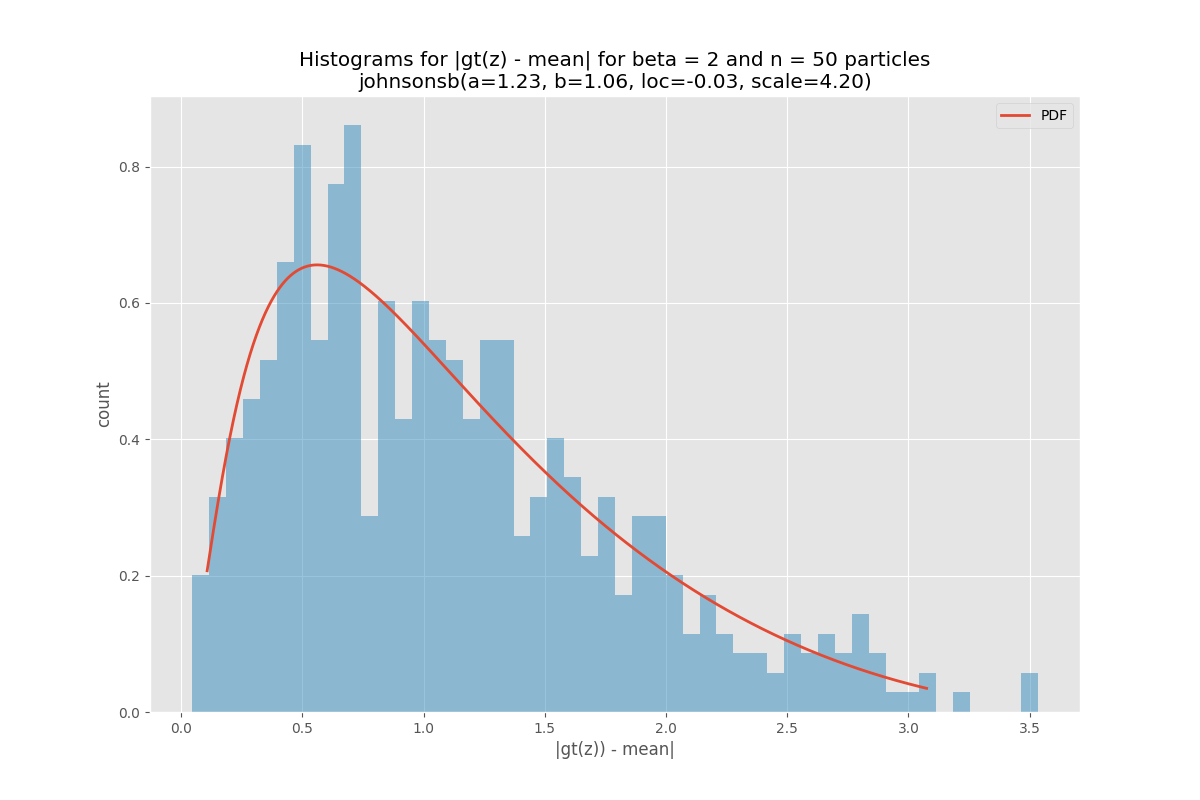
       }
       \caption{$N \cdot |g_{0.25}(1.02i) - \overline{g_{0.25}(1.02i)}|$}
    \end{subfigure}
    \caption{Histograms for $\beta = 2$ and $N = 50$.}
    \label{fig:1DBM2_50}
\end{figure}
\begin{figure}[H]
    \begin{subfigure}{0.5\textwidth}
        \centering
        \includegraphics[width=\textwidth]{
            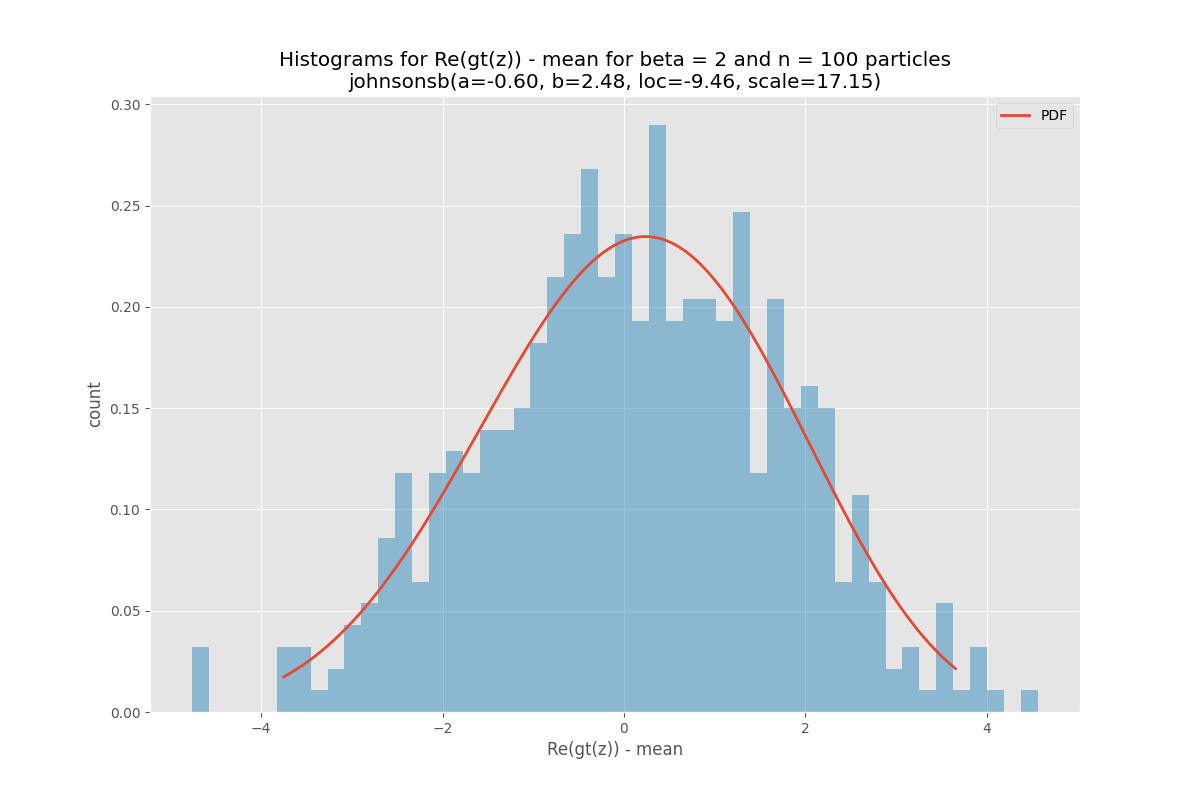
        }
        \caption{$N \cdot(Re(g_{0.25}(3i)) - \overline{Re(g_{0.25}(3i))})$}
    \end{subfigure}
    \begin{subfigure}{0.5\textwidth}
        \centering
       \includegraphics[width=\textwidth]{
           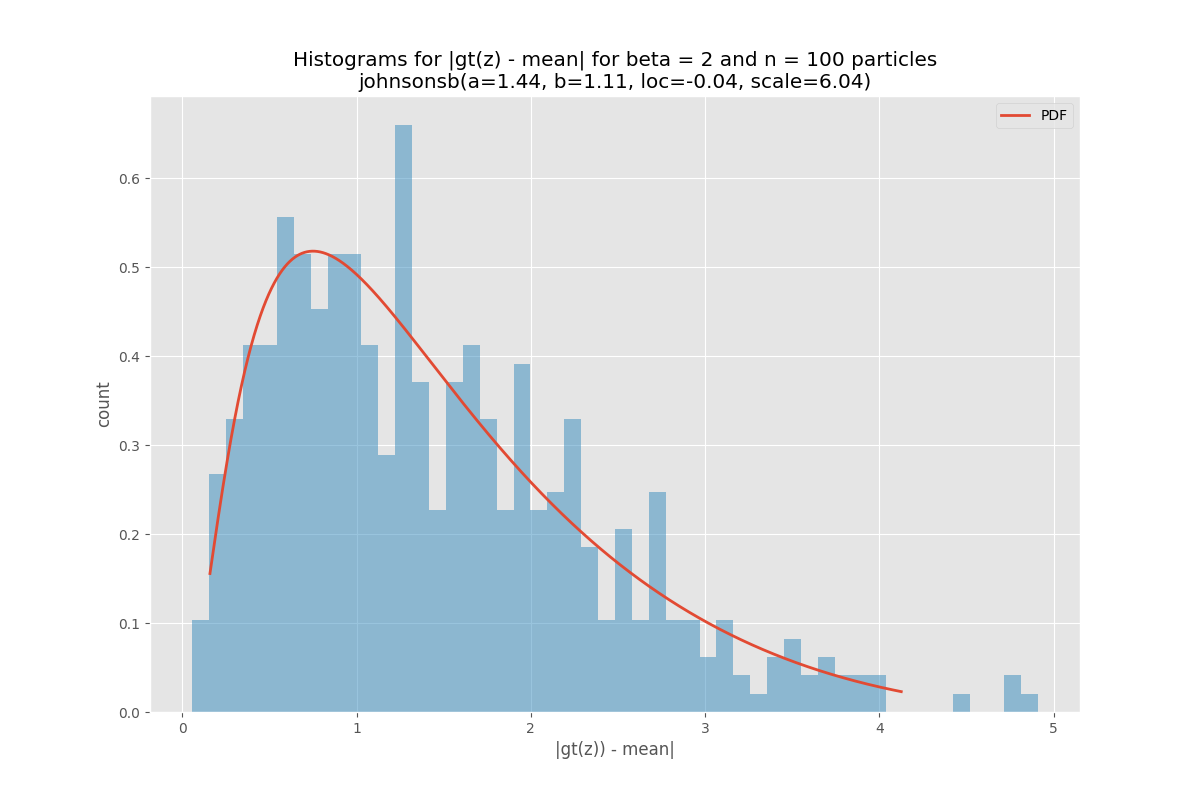
       }
       \caption{$N \cdot |g_{0.25}(1.02i) - \overline{g_{0.25}(1.02i)}|$}
    \end{subfigure}
    \caption{Histograms for $\beta = 2$ and $N = 100$.}
    \label{fig:1DBM2_100}
\end{figure}

\section{Numerical experiments studying the spread for the mulitple SLE maps dynamics started from $z_0 = 3i$} \label{app:DBM3}
\begin{figure}[H]
    \begin{subfigure}{0.5\textwidth}
        \centering
        \includegraphics[width=\textwidth]{
            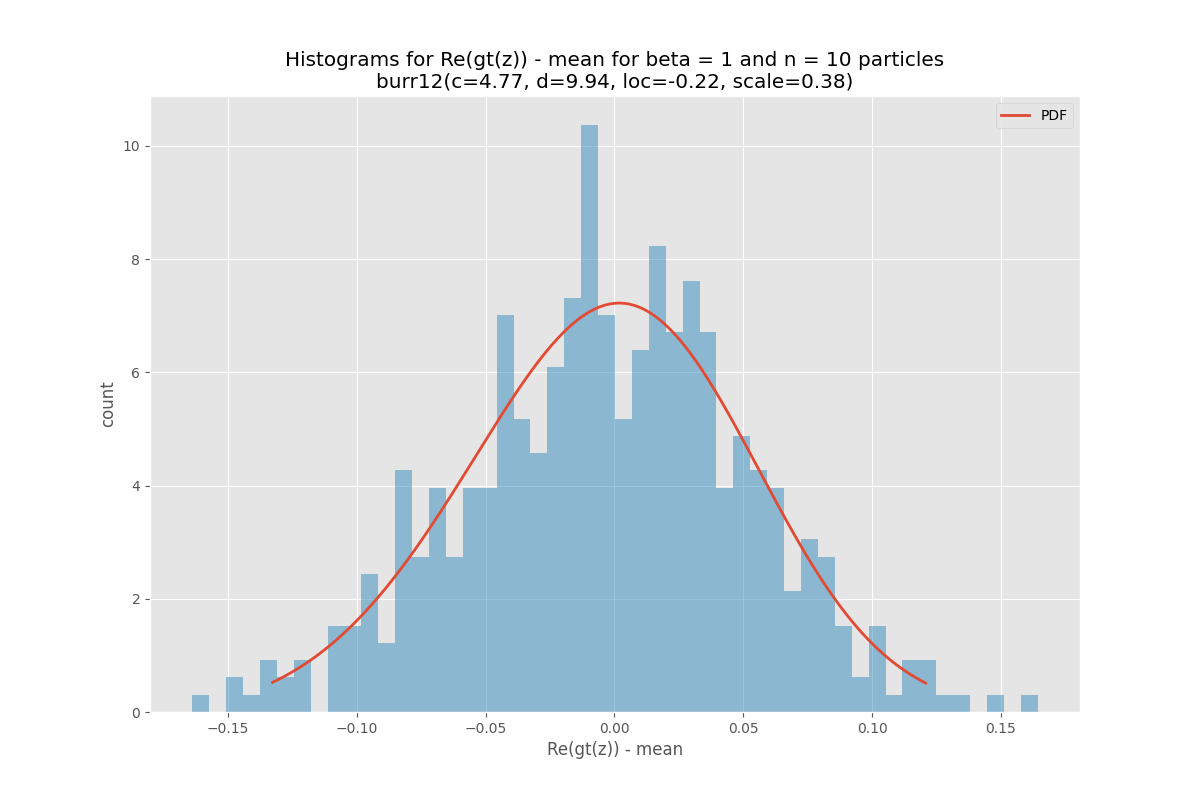
        }
        \caption{$N \cdot(Re(g_{0.25}(3i)) - \overline{Re(g_{0.25}(3i))})$}
    \end{subfigure}
    \begin{subfigure}{0.5\textwidth}
        \centering
       \includegraphics[width=\textwidth]{
           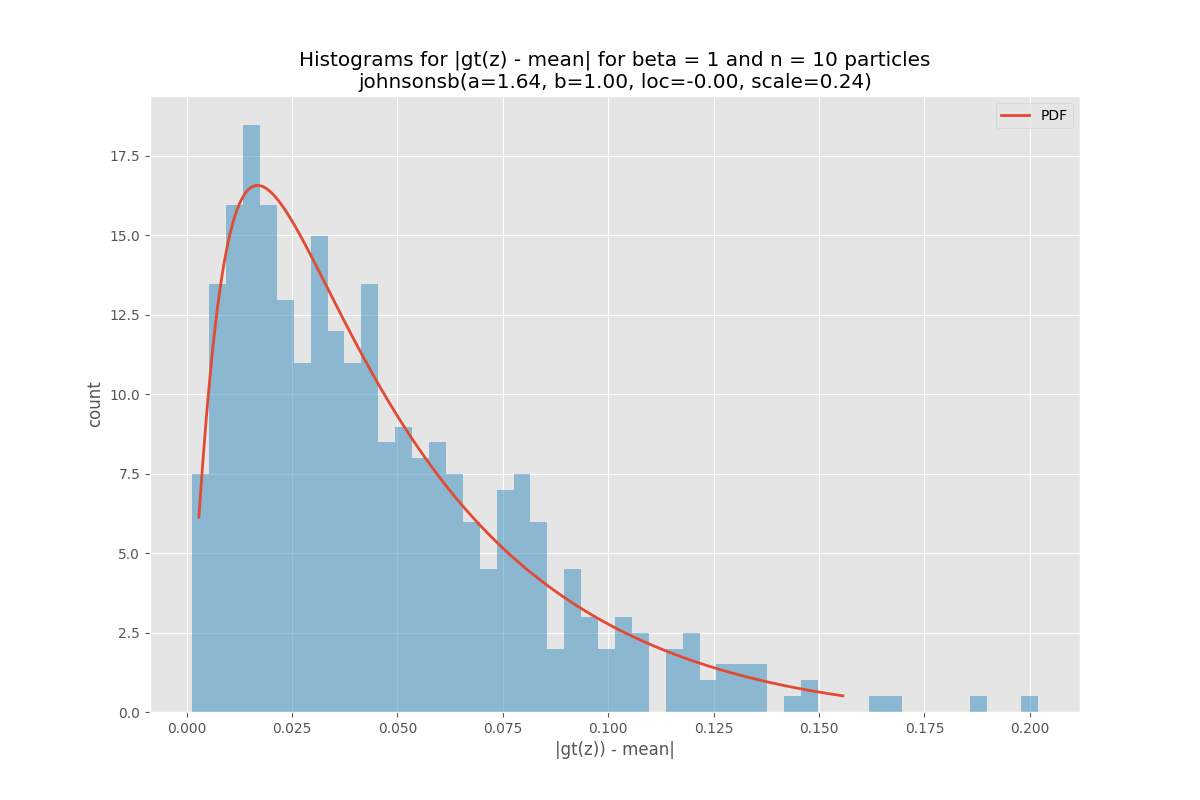
       }
       \caption{$N \cdot |g_{0.25}(3i) - \overline{g_{0.25}(3i)}|$}
    \end{subfigure}
    \caption{Histograms for $\beta = 1$ and $N = 10$.}
    \label{fig:3DBM1_10}
\end{figure}
\begin{figure}[H]
    \begin{subfigure}{0.5\textwidth}
        \centering
        \includegraphics[width=\textwidth]{
            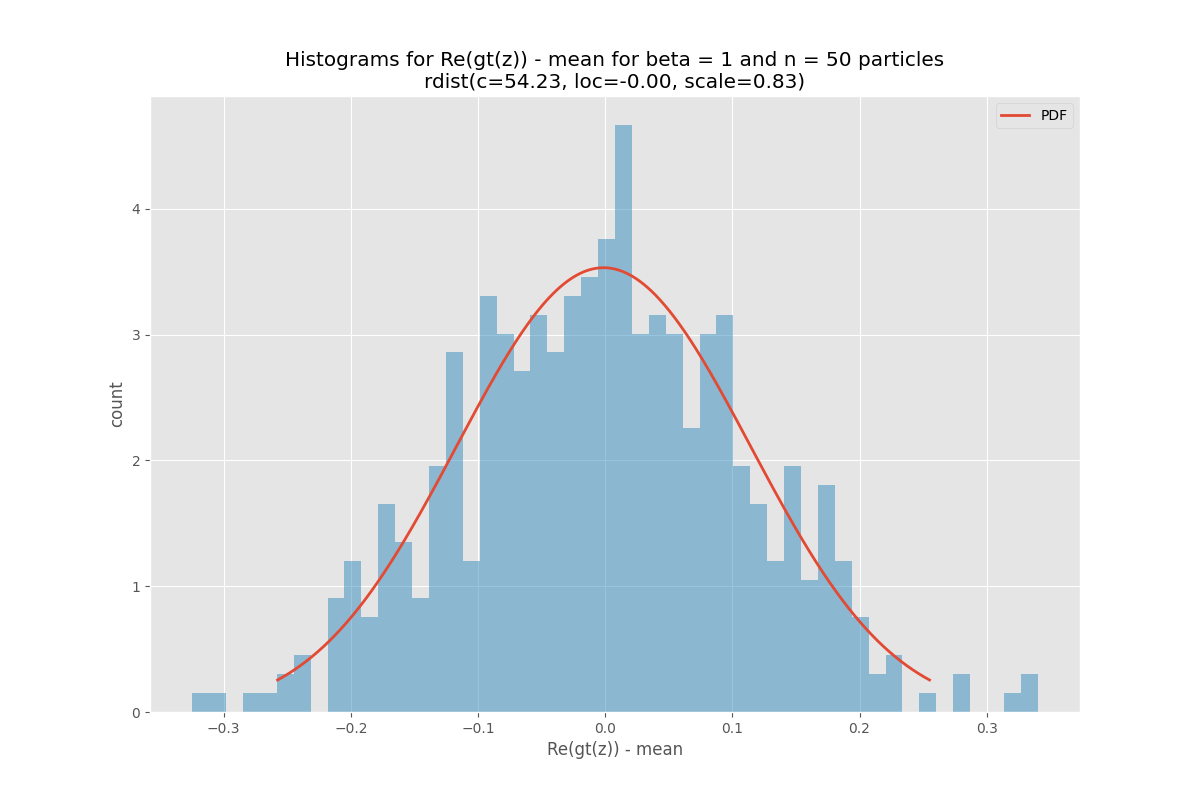
        }
        \caption{$N \cdot(Re(g_{0.25}(3i)) - \overline{Re(g_{0.25}(3i))})$}
    \end{subfigure}
    \begin{subfigure}{0.5\textwidth}
        \centering
       \includegraphics[width=\textwidth]{
           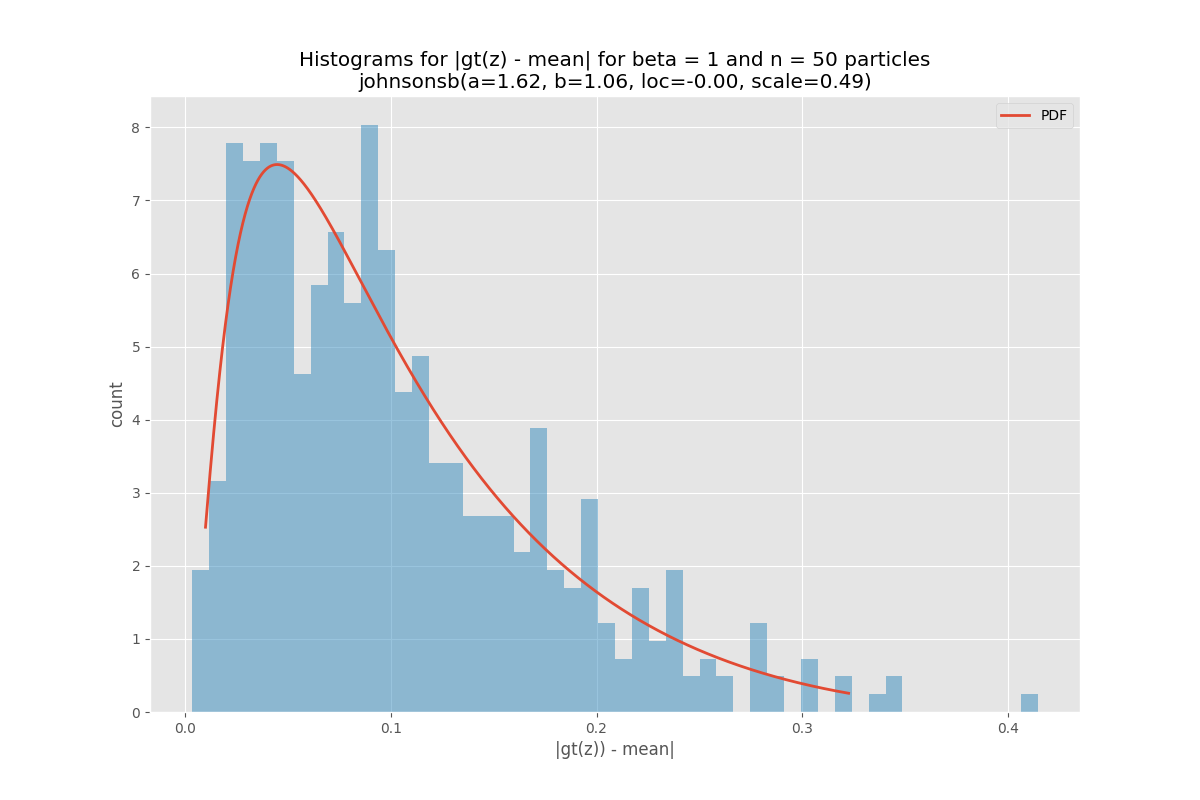
       }
       \caption{$N \cdot |g_{0.25}(3i) - \overline{g_{0.25}(3i)}|$}
    \end{subfigure}
    \caption{Histograms for $\beta = 1$ and $N = 50$.}
    \label{fig:3DBM1_50}
\end{figure}
\begin{figure}[H]
    \begin{subfigure}{0.5\textwidth}
        \centering
        \includegraphics[width=\textwidth]{
            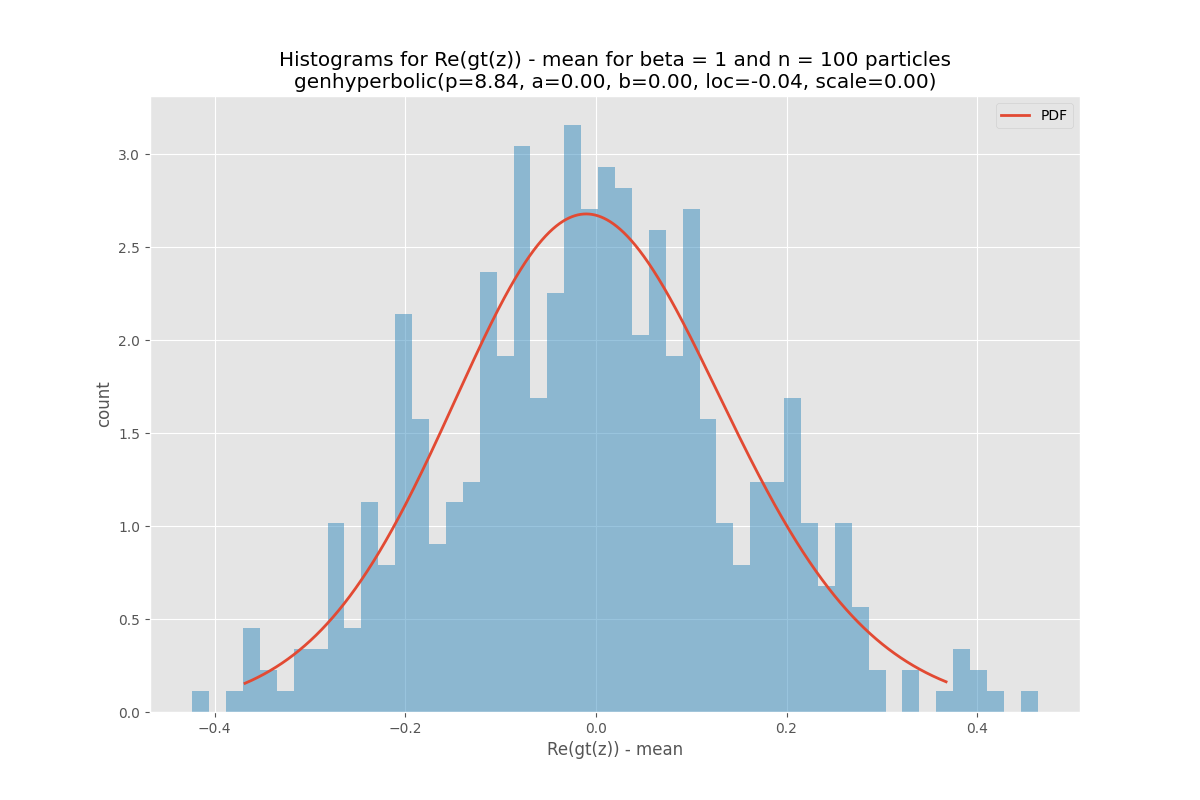
        }
        \caption{$N \cdot(Re(g_{0.25}(3i)) - \overline{Re(g_{0.25}(3i))})$}
    \end{subfigure}
    \begin{subfigure}{0.5\textwidth}
        \centering
       \includegraphics[width=\textwidth]{
           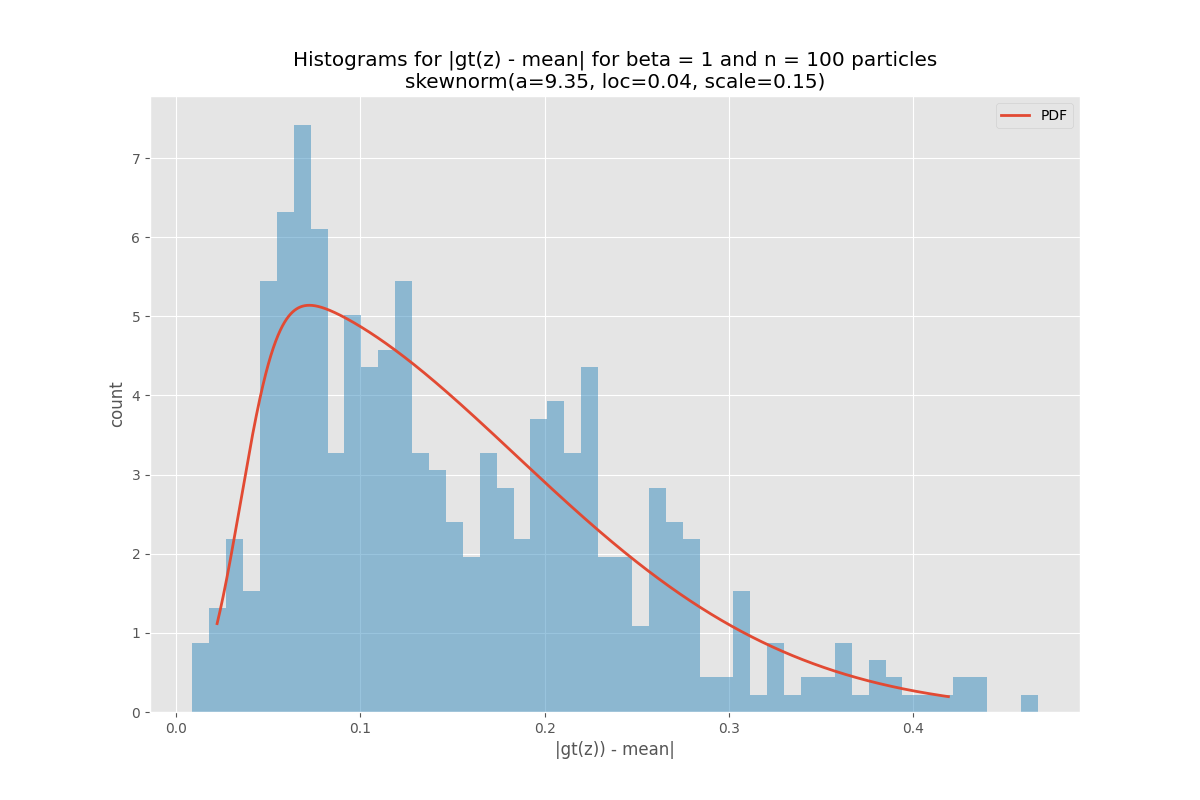
       }
       \caption{$N \cdot |g_{0.25}(3i) - \overline{g_{0.25}(3i)}|$}
    \end{subfigure}
    \caption{Histograms for $\beta = 1$ and $N = 100$.}
    \label{fig:3DBM1_100}
\end{figure}
\begin{figure}[H]
    \begin{subfigure}{0.5\textwidth}
        \centering
        \includegraphics[width=\textwidth]{
            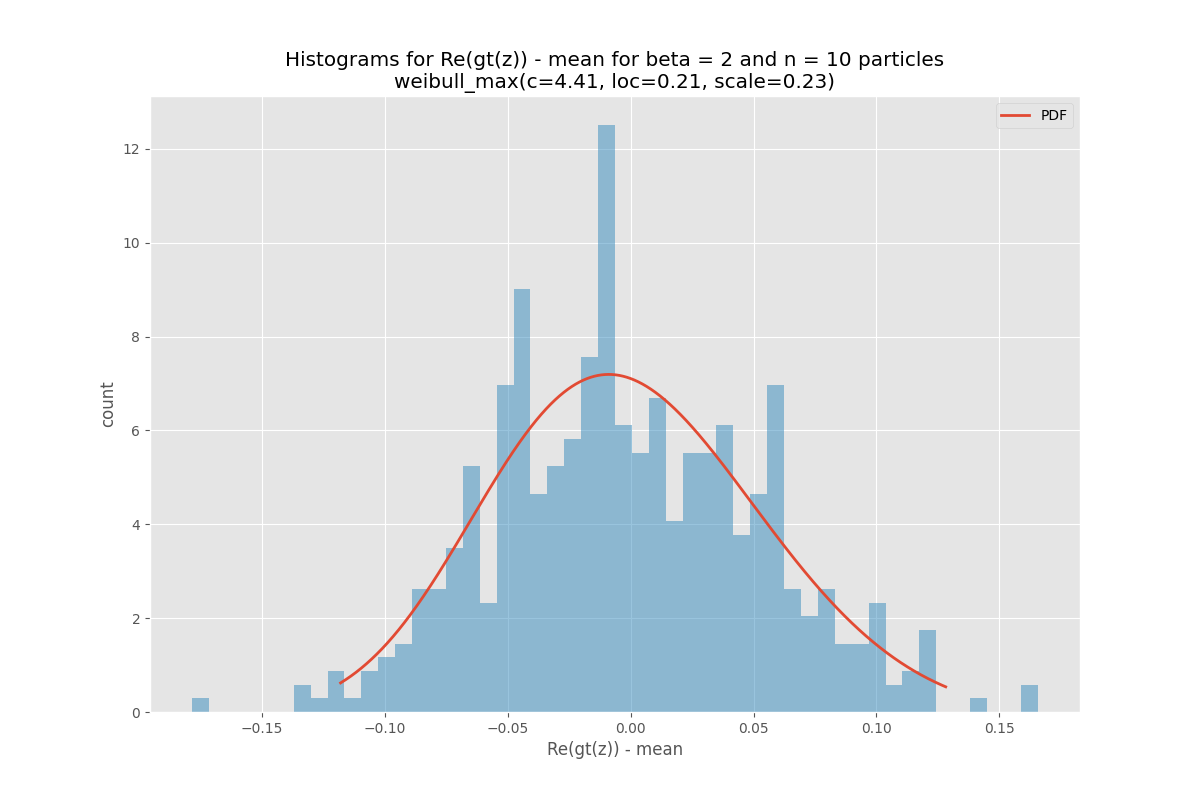
        }
        \caption{$N \cdot(Re(g_{0.25}(3i)) - \overline{Re(g_{0.25}(3i))})$}
    \end{subfigure}
    \begin{subfigure}{0.5\textwidth}
        \centering
       \includegraphics[width=\textwidth]{
           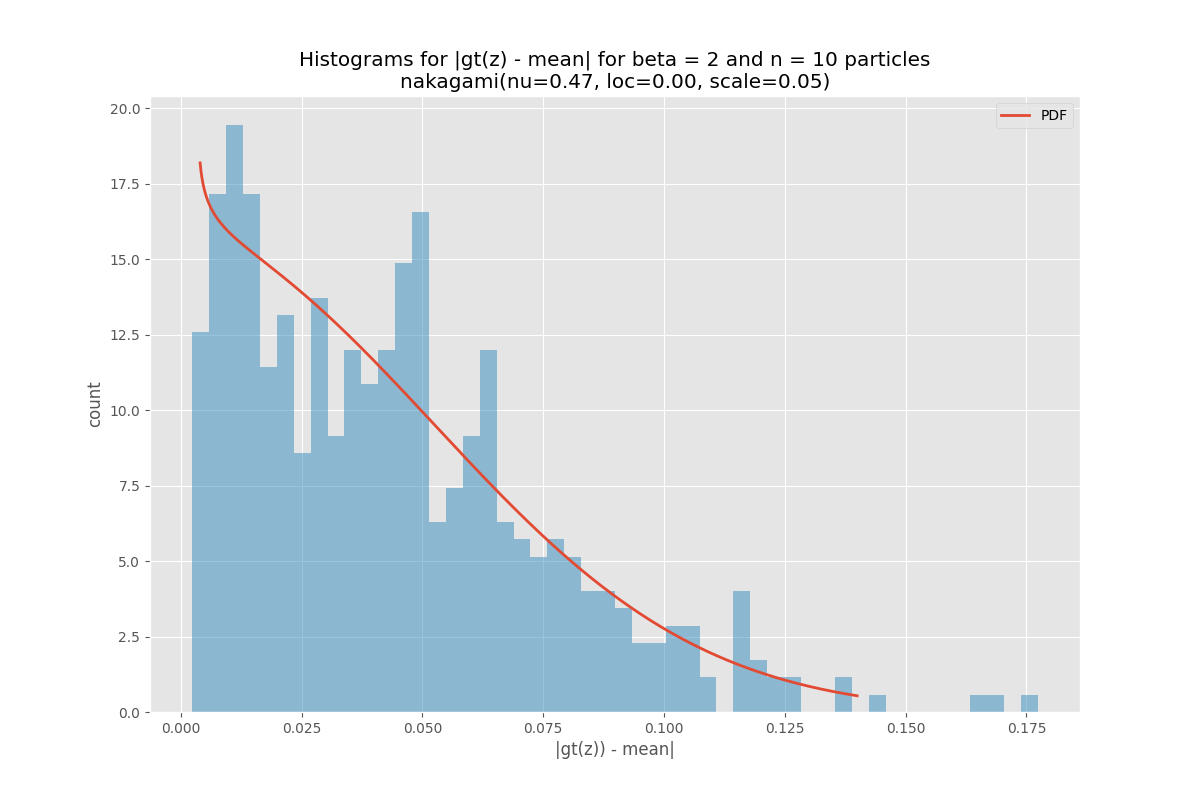
       }
       \caption{$N \cdot |g_{0.25}(3i) - \overline{g_{0.25}(3i)}|$}
    \end{subfigure}
    \caption{Histograms for $\beta = 2$ and $N = 10$.}
    \label{fig:3DBM2_10}
\end{figure}
\begin{figure}[H]
    \begin{subfigure}{0.5\textwidth}
        \centering
        \includegraphics[width=\textwidth]{
            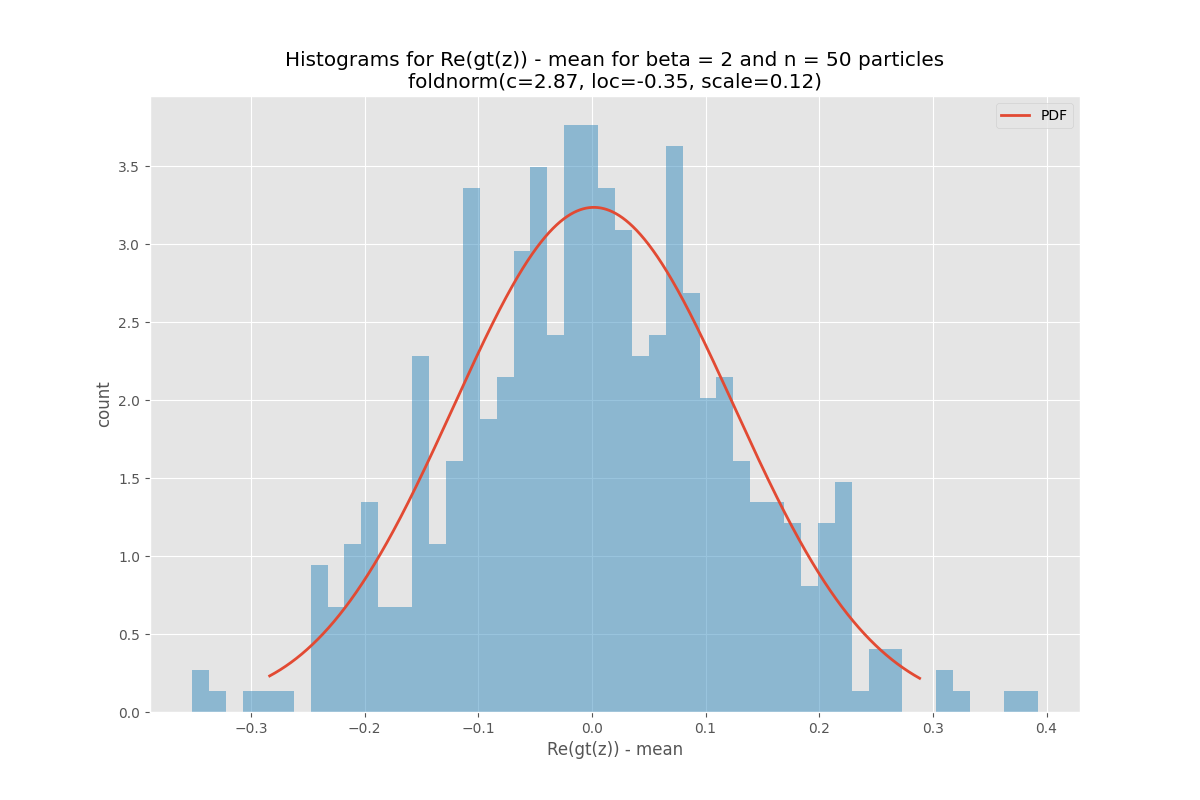
        }
        \caption{$N \cdot(Re(g_{0.25}(3i)) - \overline{Re(g_{0.25}(3i))})$}
    \end{subfigure}
    \begin{subfigure}{0.5\textwidth}
        \centering
       \includegraphics[width=\textwidth]{
           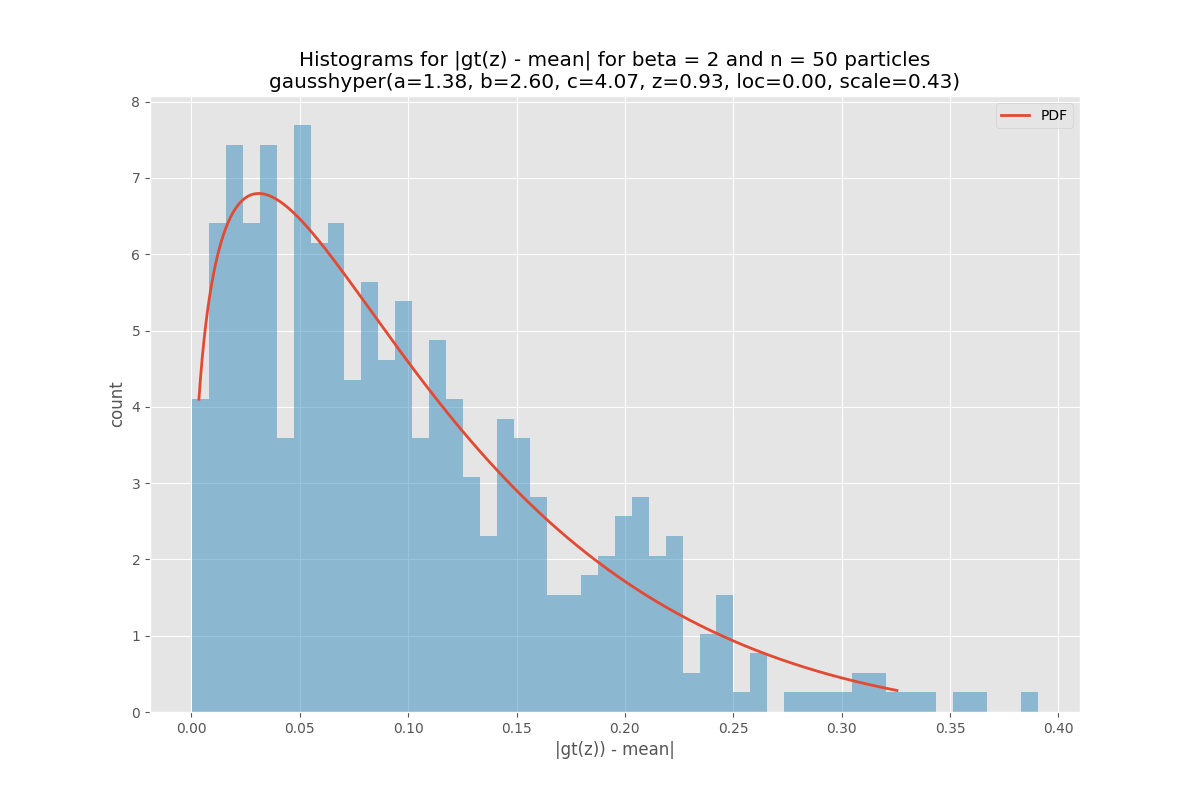
       }
       \caption{$N \cdot |g_{0.25}(3i) - \overline{g_{0.25}(3i)}|$}
    \end{subfigure}
    \caption{Histograms for $\beta = 2$ and $N = 50$.}
    \label{fig:3DBM2_50}
\end{figure}
\begin{figure}[H]
    \begin{subfigure}{0.5\textwidth}
        \centering
        \includegraphics[width=\textwidth]{
            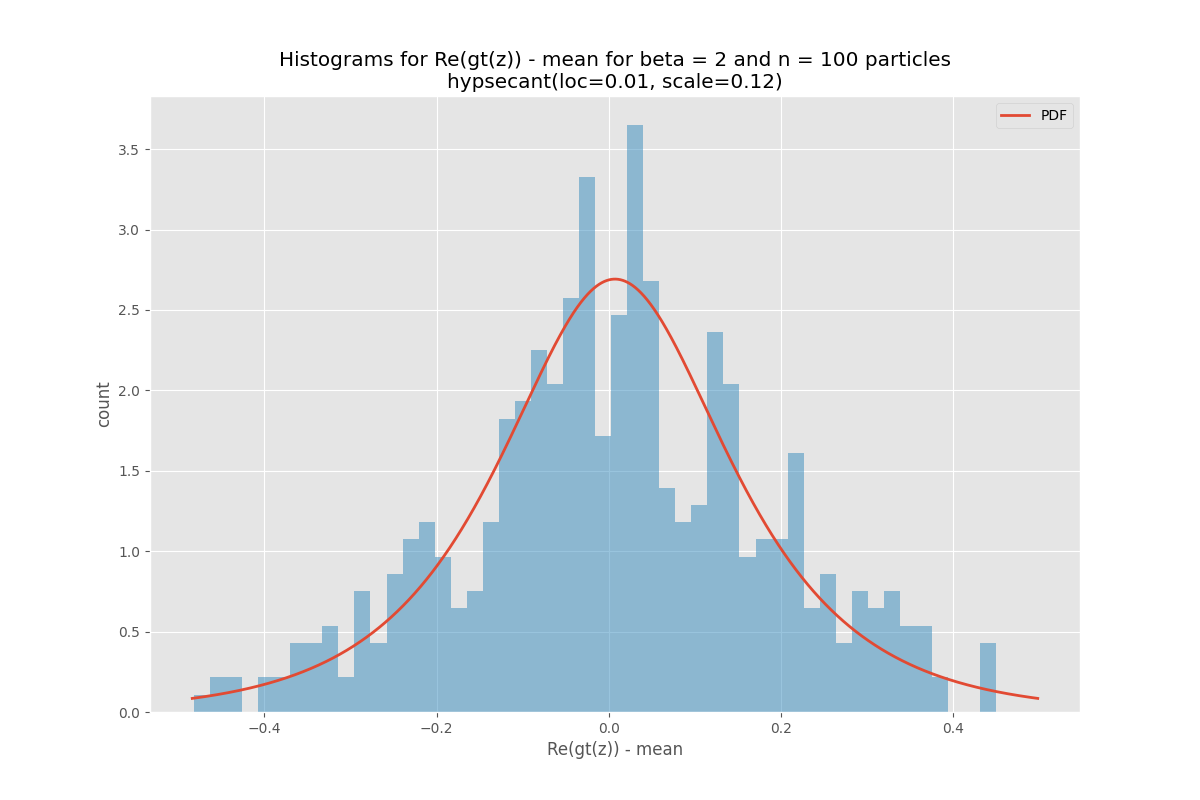
        }
        \caption{$N \cdot(Re(g_{0.25}(3i)) - \overline{Re(g_{0.25}(3i))})$}
    \end{subfigure}
    \begin{subfigure}{0.5\textwidth}
        \centering
       \includegraphics[width=\textwidth]{
           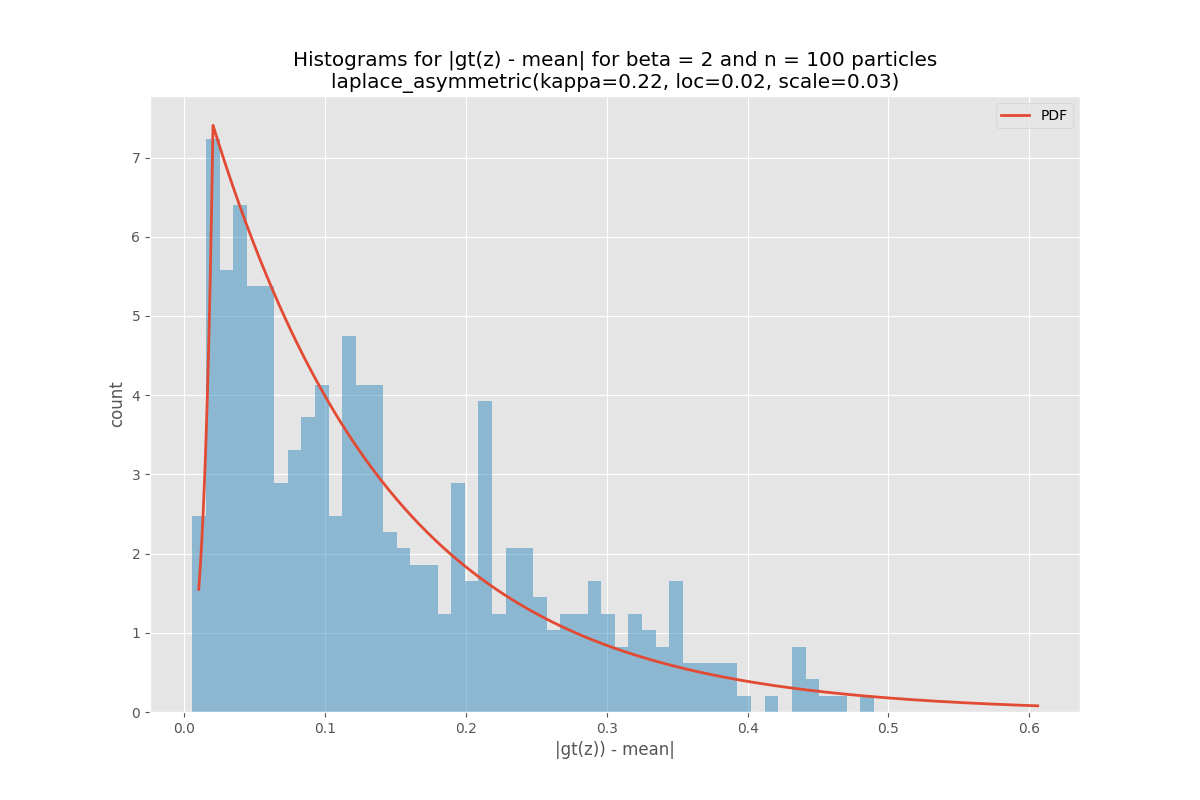
       }
       \caption{$N \cdot |g_{0.25}(3i) - \overline{g_{0.25}(3i)}|$}
    \end{subfigure}
    \caption{Histograms for $\beta = 2$ and $N = 100$.}
    \label{fig:3DBM2_100}
\end{figure}
\end{document}